\documentclass[aps,a4paper,amssymb,amsmath,reprint,superscriptaddress]{revtex4-1}

\usepackage[dvipsnames]{xcolor}
\usepackage{amsmath,amssymb,bm,braket,ulem}

\usepackage[hiresbb,pdftex]{graphicx} 
\usepackage[
pdftex,
draft=false,
bookmarks=true,
bookmarksnumbered=false,
bookmarksopen=false,
bookmarksopenlevel=4,
colorlinks=true,
anchorcolor=black,
citecolor=blue,
filecolor=black,
linkcolor=black,
linkbordercolor={1 0 0},
urlcolor=violet,
pdfborder={0 0 1},
pdftitle={},
pdfauthor={},
pdfsubject={},
pdfkeywords={},
pdfpagemode=UseThumbs,
]{hyperref}

\newcommand{\Pa}{$\mathcal{P}$}
\newcommand{\T}{$\mathcal{T}$}
\newcommand{\PT}{$\mathcal{PT}$}

\newcommand{\limvec}{\lim_{\bm{\lambda}\rightarrow \bm{0}}}
\newcommand{\vj}{{J}}
\newcommand{\bk}{{\bm{k}}}
\newcommand{\bA}{{\bm{A}}}
\newcommand{\bla}{{\bm{\lambda}}}
\newcommand{\cobc}{\mathcal{A}}
\newcommand{\cm}{$\checkmark$}

\newcommand{\LP}{$\updownarrow$}
\newcommand{\CP}{$\circlearrowleft$}

\newcommand{\mr}[2]{#1 \, {\mathrm{ \,  #2 \,}}}
\DeclareMathOperator{\Tr}{Tr}

\begin{document} 

\title{Nonreciprocal optical response in parity-breaking superconductors}
\author{Hikaru Watanabe}
\affiliation{RIKEN Center for Emergent Matter Science (CEMS), Wako 351-0198, Japan}
\affiliation{Department of Physics, Graduate School of Science, Kyoto University, Kyoto 606-8502, Japan}
\author{Akito Daido}
\affiliation{Department of Physics, Graduate School of Science, Kyoto University, Kyoto 606-8502, Japan}
\author{Youichi Yanase}
\affiliation{Department of Physics, Graduate School of Science, Kyoto University, Kyoto 606-8502, Japan}
\affiliation{Institute for Molecular Science, Okazaki 444-8585, Japan}
\date{\today}

\begin{abstract}
    Superconductivity, an emergence of a macroscopic quantum state, gives rise to unique electromagnetic responses leading to perfect shielding of magnetic field and zero electrical resistance.
    In this paper, we propose that superconductors with the space-inversion symmetry breaking host giant nonreciprocal optical phenomena, such as photocurrent generation and second harmonic generation.
    The nonreciprocal optical responses show divergent behaviors in the low-frequency regime and originate from two-fold indicators unique to parity-breaking superconductors, namely, the nonreciprocal superfluid density and the Berry curvature derivative.
    Furthermore, the mechanism and frequency dependence are closely tied to the preserved temporal symmetry in the superconductor. The relation is useful for probing the space-time structure of the superconducting symmetry.
    The indicators characterize the low-frequency property of nonreciprocal optical responses and hence quantify the performance of superconductors in nonreciprocal optics.
    \end{abstract}
\maketitle

\section{Introduction}

Nonlinear optics is one of the central fields in physics and has been offering extensive interests from fundamental science to engineering.
When the system has no space-inversion (\Pa{}) symmetry, the nonlinear optical response acquires the nonreciprocal property, that is, asymmetry in the response to mutually antiparallel external fields.
A prototypical example of the nonreciprocal response is the diode effect where the electric conductivity is allowed or almost forbidden depending on the direction of voltage.

The nonreciprocal optical (NRO) phenomena have been explored in various systems~\cite{Fiebig2005,Orenstein2021}. The leading NRO response is represented by the second-order nonlinear response. The formula is given by the current response induced by the double photo-electric field
    \begin{align}
    \Braket{\mathcal{J}^\alpha_\text{NRO}  (\omega)}
            = \int \frac{d\Omega}{2\pi}\sigma^{\alpha;\beta\lambda}\, (\omega;\Omega,\omega-\Omega) E^\beta (\Omega) E^\lambda (\omega- \Omega).\label{NRO_response_formula}
    \end{align}
The NRO response covers two important phenomena, namely, second harmonic and photocurrent generations which are denoted by $\sigma (2\omega;\omega,\omega)$ and $\sigma (0;\omega,-\omega)$, respectively.

The second harmonic generation, where irradiating light with frequency $\omega$ induces the light oscillating with the doubled frequency $2\omega$, is a useful tool for probing the microscopic \Pa{} breaking in materials.
In the field of multiferroic material science, the second harmonic generation led to the successful optical imaging of magnetic domains~\cite{Fiebig2005}.
Furthermore, recent second harmonic generation experiments implied exotic parity-violating ordering in quantum materials such as a family of high-temperature copper-based superconductors~\cite{Zhao2017,De_la_Torre2020}. 

On the one hand, the photo-induced direct current, namely, the photocurrent generation (photogalvanic effect), is also of interest.
The conventional mechanism for the photocurrent response is attributed to spatially-inhomogeneous and asymmetric structures found in photo-diode devices,  whereas recent studies shed light on the bulk photocurrent response originating from microscopic parity breaking~\cite{Sturman1992Book}.
Since the bulk photocurrent performance is influenced by the electronic property of materials, the realization of a novel type of photo-electric converter is one of the strong motivations for materials science.
Potential candidate systems are topological materials~\cite{Liu2020SemimetalReview,Orenstein2021} and parity-violating magnets~\cite{Ogawa2016,Burger2020}.
In the former systems, the photocurrent response is enhanced by the nontrivial quantum geometry in the electronic band structure, while the latter systems enable a tunable photocurrent response due to the good controllability of magnetic order.

    \begin{figure*}[t]
    \centering
    \includegraphics[width=0.90\linewidth,clip]{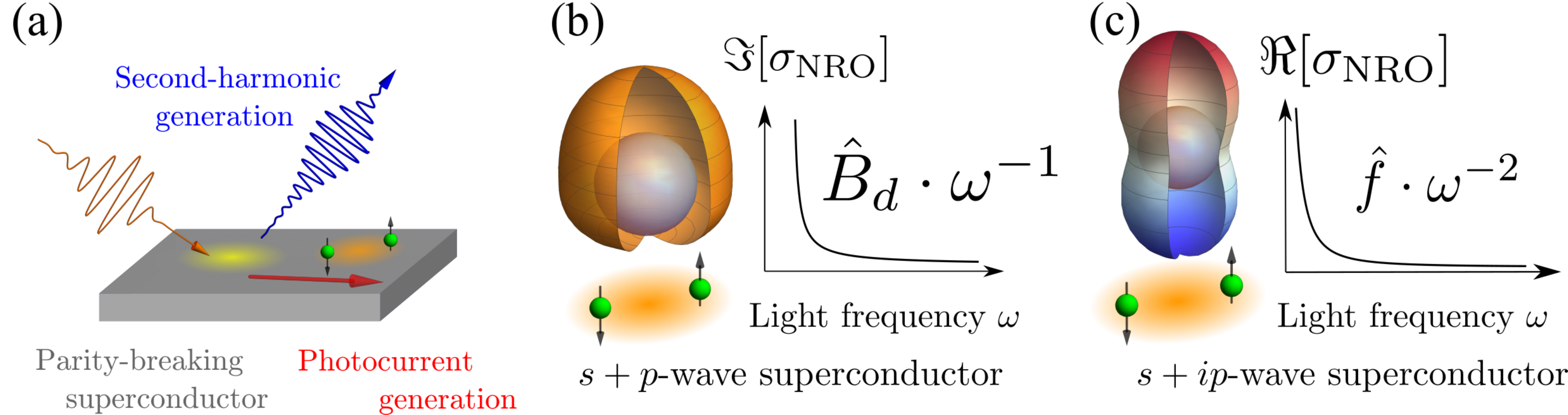}
    \caption{
    Nonreciprocal optics in superconductors.
    (a) Sketch of nonreciprocal optical responses in a parity-breaking superconductor. The Cooper pairs (paired electrons) are colored in green.
    The injected light generates light with doubled frequency (second harmonic generation) and direct current (photocurrent generation).
    (b) $s+p$-wave superconductivity is an example of a time-reversal symmetric parity-breaking superconductor.
    Its characteristic NRO response is the diverging imaginary component of the NRO conductivity ($\Im{[\,\sigma_\text{NRO}]}$), which is determined by the Berry curvature derivative $\hat{B_d}$.
    (c) $s+ip$-wave superconductivity is a parity-time-reversal symmetric parity-breaking superconductor.
    The nonreciprocal superfluid density $\hat{f}$ contributes to the diverging NRO conductivity in the real part ($\Re{[\,\sigma_\text{NRO}]}$).
    }
    \label{Fig_schematics}
    \end{figure*}

Stimulated by the progress mentioned above, we investigate the NRO responses in another class of quantum materials, \textit{superconductors}.
Superconductors host striking electromagnetic properties stemming from the Cooper pairs' quantum condensation.
Well-known examples are the perfect shielding of magnetic flux (Meissner effect) and zero resistivity phenomenon.
The optical conductivity also shows a unique spectrum in the low-frequency region and has been exploited to evaluate the superfluidity of Cooper pairs~\cite{Tinkham2004introduction}.
The nontrivial quantum nature of superconductors is also attracting from the viewpoint of nonreciprocal optics.
The dissipation-free electric conductivity of superconductors is favorable for suppressing undesirable energy loss in the photo-electric conversion process.
It is therefore expected that superconductivity paves a new way for optoelectronics.
Supporting this prospect, enhanced nonreciprocal direct conductivity has been reported in superconductors~\cite{Tokura2018,Ideue2020} and the ``perfect'' diode effect has recently been observed in a parity-breaking superconductor~\cite{Ando2020}.
Despite the potential applications to these transport responses, only a few works were conducted for exploring NRO responses of superconductors~\cite{Zhao2017,Xu2019,Yang2019,Nakamura2020,Vaswani2020,Lim2020}.
In particular, unique features of the NRO response in superconductors remain to be elusive.
To motivate further interest in superconductor-based optoelectronics, in this paper, we predict a giant NRO response of parity-violating superconductors.

The key features of the NRO response are two-fold.
One is the absence of the definite \Pa{} parity in the superconducting state.
The other is the temporal symmetry, such as the time-reversal (\T{}) symmetry and the symmetry of the combined \Pa{} and \T{} operations, called the parity-time-reversal (\PT{}) symmetry.
Based on the symmetry analysis and microscopic calculations, in this paper, we demonstrate that the parity-breaking superconductor shows the diverging NRO response in the low-frequency regime.
This newly discovered response is useful for sensitive photo-detection and energy harvesting.
We also show that the divergent behavior shows the characteristic power law associated with two indicators, namely, the \textit{Berry curvature derivative} $\hat{B_d}$ and the \textit{nonreciprocal superfluid density} $\hat{f}$, which are closely tied to the temporal symmetry.
Figure~\ref{Fig_schematics} overviews this work with a typical example of parity-breaking superconductors comprising the even-parity $s$-wave and odd-parity $p$-wave pairings.

The outline of the paper is given below.
In Sec.~\ref{Sec_symmetry_analysis}, we present the symmetry analysis to identify superconductors offering the NRO responses and classify potential candidates.
Section~\ref{Sec_formulation} briefly explains the formulation of the NRO response in superconductors. Starting from the general expression obtained with the Bogoloubov-de Gennes formalism, we propose the superconductivity-induced NRO response, which we call the anomalous NRO response.
The analytical formulas are in agreement with those obtained by two-fold methods, namely, density matrix and Green function methods (the derivations are detailed in Appendices~\ref{App_Sec_derivation_photocurrent_response} and~\ref{App_Sec_Green_function_method_derivation}).
The results are summarized in Table~\ref{Table_photocurrent_classification}.
Furthermore, making use of the adiabatic time-evolution process, we obtain formulas for the anomalous nonreciprocal responses which can apply to many-body Hamiltonians (Sec.~\ref{Sec_manybody}).
The analytical results are corroborated by the numerical calculations of the NRO responses such as photocurrent and second harmonic generations in Sec.~\ref{Sec_model_study}.
The contents are concluded in Sec.~\ref{Sec_summary}.

Throughout this paper, we present formulas with the unit $\hbar=1$ (Dirac constant), $q=1$ (electron charge).

\section{Symmetry and candidate superconductors}
\label{Sec_symmetry_analysis}

First, we present the symmetry analysis of candidate superconductors for the nonreciprocal optics.
Broken space-inversion symmetry is an essential ingredient for the NRO response in the superconducting state as well.
On the other hand, the emergence of superconductivity gives the stiffness of the U(1)-gauge transformation.
Therefore, it is relevant to take into account the combined operations of the usual symmetry operations and U(1)-gauge transformation.
In the case of the odd-parity superconductors, the combined inversion operation \Pa{}$\times$U(1) is still preserved although the usual \Pa{} symmetry is broken.
The surviving ``inversion'' symmetry also prohibits the NRO conductivity.
Thus, we cannot expect any NRO responses in purely odd-parity superconductors, \textit{e.g.}, centrosymmetric ferromagnetic superconductors such as UGe$_2$, URhGe, and UCoGe~\cite{Aoki_review2019}.
Multiple superconducting transitions in UPt$_3$ are also expected to be caused by purely odd-parity pairings~\cite{Sauls_review1994,Joynt_review2002}.
This symmetry requirement is different from that of the diagonal long-range order invoking parity violation such as structural transitions and magnetic orderings~\cite{Watanabe2021}.

According to the above-mentioned symmetry considerations, superconductors hosting the NRO response have no definite parity under the \Pa{} operation.
To our best knowledge, the parity violation realizes in the following setups; (I) superconductivity occurs in the system where \Pa{} symmetry has already been broken in the normal state, (II) an external field induces parity breaking unique to superconductors, and (III) multiple superconducting transitions are caused by coexisting even-parity and odd-parity Cooper pairings.

Case (I) may exist in a broad range of materials since there are a series of noncentrosymmetric bulk superconductors such as CePt$_3$Si and Li$_2$(Pd,Pt)$_3$B~\cite{NCSC_book}.
In addition, parity-breaking effects have recently been demonstrated~\cite{Saito2016,Lu2015,STO_parity} in artificial two-dimensional superconductors such as ultrathin MoS$_2$~\cite{Ye2012} and SrTiO$_3$/LaAlO$_3$ interfaces~\cite{Reyren2007}.
Other situations with broken \Pa{} symmetry can be found when superconductivity coexists with the structural transition~\cite{Hiroi2018} and magnetic ordering~\cite{Zhao2017,Sumita2017}.

Corresponding to Case (II), an electric current breaks the \Pa{} and \T{} symmetry. 
Different from the metal state, the superconducting state allows the electric current to flow in thermal equilibrium with the aid of the superfluidity.
Under the electric current, Cooper pairs acquire finite momentum and the Fulde-Ferrell superconductivity is realized~\cite{FF1964}, which is classified as the \PT{} symmetric parity-breaking superconducting state.
Thus, it is expected that superconductors exhibit NRO responses under the supercurrent flow.
The experimental observations have been reported in Refs.~\cite{Yang2019,Nakamura2020,Vaswani2020}.

Finally, we consider Case (III).
This class of the parity-breaking superconductors is intrinsic, nontrivial, and exotic because the superconductivity itself breaks both of the \Pa{} symmetry and its combination with the U(1) gauge transformation.
The resulting parity violation is solely induced by superconducting instability, and thus it is in sharp contrast to the known parity-breaking superconductors classified into Cases (I) and (II).
The spontaneously parity-mixed superconductivity may be stabilized near the topological phase transition between $s$-wave and $p$-wave superconducting states~\cite{Wang2017}.
It was shown that the relative phase of the even and odd-parity pair potentials is $\pm \pi/2$ in a coexisting phase space.
The resulting parity-mixed superconductivity preserves \PT{} symmetry.
Although this scenario requires a tunable parameter for the topological phase transition, the recent discovery of the heavy fermion superconductor UTe$_2$~\cite{Ran2019}, which shows multiple superconducting transitions~\cite{ran2019extreme,Braithwaite2019,Aoki2020,Knebel2020,Lin2020,aoki2021fieldinduced},
led to a theoretical prediction of the spontaneously parity-mixed superconductivity~\cite{Ishizuka2021}.

Other superconductivity-driven symmetry breaking was also studied in terms of chiral superconductivity~\cite{Kallin_2016,Brydon2019} and nematic superconductivity~\cite{Yonezawa2019}.
The time-reversal symmetry breaking has been discussed based on the $\mu$SR and Kerr rotation measurements~\cite{Kallin_2016}, while the rotation symmetry breaking has been studied by various bulk measurements~\cite{Yonezawa2019}. 
However, experimental probes for detecting intrinsic space-inversion symmetry breaking are awaited to verify a new phase of matter. 
The NRO response may be useful for detecting parity violation induced by superconductivity and for evaluating its parity-mixing effect.
In the numerical calculations, we demonstrate the NRO responses arising from the parity violation of Case (III) as an example, while our formulation applies to all the cases (I)-(III).

\section{Formulation of NRO responses in superconductors}
\label{Sec_formulation}

\subsection{General formula for NRO responses}
\label{Secsub_setup}

Our formulation is based on the Bogoliubov-de Gennes (BdG) Hamiltonian 
        \begin{equation}
        \mathcal{H}_\text{BdG} = \frac{1}{2} \bm{\Psi}^\dagger H_\text{BdG}\bm{\Psi} +\text{const.},\label{eq:BdGtemp}
        \end{equation}
where Nambu spinor $\bm{\Psi}^\dagger = (\bm{c}^\dagger,\bm{c}^T)$ with the creation ($\bm{c}^\dagger$) and annihilation ($\bm{c}$) operators of electrons.
The BdG Hamiltonian $H_{\text{BdG}}$ consists of the normal-state Hamiltonian $H_\text{N}$ and the pair potential $\Delta$
        \begin{equation}
            H_{\text{BdG}}(\bm{A})
                =\begin{pmatrix}
                    H_\text{N}(\bm{A})&\Delta\\
                    \Delta^\dagger&-[H_\text{N}(\bm{A})]^T
                \end{pmatrix}.
            \label{eq:BdG_A}
        \end{equation}
The vector potential $\bA$ dependence is taken into account for the electromagnetic field perturbation, while we choose $\bA=\bm{0}$ in the equilibrium. Following the standard perturbative treatments (Appendix~\ref{App_Sec_density_matrix_formulation}), we evaluate the expectation value of the electric current density as
		\begin{equation}
            \Braket{\mathcal{J}^\alpha  (\omega)} = \sum_{n=1} \Braket{\mathcal{J}^\alpha (\omega)}_{(n)},
        \end{equation}
where $\Braket{\mathcal{J}^\alpha (\omega)}_{(n)}$ is the electric current of the $n$-th order in $\bm{A}$.
With the velocity gauge $\bm{E} = -\partial_t \bm{A} (t)$, we obtain the NRO response function in Eq.~\eqref{NRO_response_formula} corresponding to the second-order component $\Braket{\mathcal{J}^\alpha (\omega)}_{(n=2)}$.
The formula is given by
        \begin{align}
            & \sigma^{\alpha;\beta\gamma} (\omega;\omega_1,\omega_2) \nonumber \\
                & =\frac{1}{2 \left( i\omega_1-\eta \right) \left( i\omega_2-\eta \right)} \left[ \sum_a \frac{1}{2}  \vj^{\alpha\beta\gamma}_{aa} f_a \label{RDM_2;0} \right.\\ 
                &\left.+\sum_{a,b} \frac{1}{2} \left(  \frac{\vj^{\alpha\beta}_{ab} \vj^\gamma_{ba} f_{ab} }{ \omega_2 +i\eta -E_{ba}} +  \frac{\vj^{\alpha\gamma}_{ab} \vj^\beta_{ba} f_{ab} }{ \omega_1 +i\eta -E_{ba}}\label{RDM_1;1}\right)\right. \\
                &\left.+\sum_{a,b}\frac{1}{2}  \frac{\vj^{\alpha}_{ab} \vj^{\beta\gamma}_{ba} f_{ab} }{ \omega + 2i\eta -E_{ba}}\label{RDM_0;2_2photon}\right.\\
                &\left.+\sum_{a,b,c} \frac{1}{2}\frac{\vj^{\alpha}_{ab}}{ \omega + 2i\eta -E_{ba}} \left(   \frac{\vj^\beta_{bc}\vj^{\gamma}_{ca}f_{ac} }{\omega_2 + i \eta -E_{ca} }    -    \frac{\vj^\beta_{ca}\vj^{\gamma}_{bc}f_{cb} }{\omega_2 + i \eta -E_{bc} }  \right)\label{RDM_0;2_1photon_1}\right.\\
                &\left.+\sum_{a,b,c}\frac{1}{2}\frac{\vj^{\alpha}_{ab}}{ \omega + 2i\eta -E_{ba}} \left(  \frac{\vj^\gamma_{bc}\vj^{\beta}_{ca}f_{ac} }{\omega_1 + i \eta -E_{ca} }    -    \frac{\vj^\gamma_{ca}\vj^{\beta}_{bc}f_{cb} }{\omega_1 + i \eta -E_{bc} } \right)\right], \label{RDM_0;2_1photon_2}
        \end{align} 
where indices $a,b,c$ are spanned by the energy eigenstates of the unperturbed BdG Hamiltonian $H_\text{BdG} (\bm{0})$ and the eigenenergy difference is $E_{ab} = E_a -E_b$.
We introduced the Fermi-Dirac distribution function $f_a = (e^{\beta E_a}+1)^{-1}$ and accordingly defined $f_{ab} = f_a -f_b$.
The infinitesimal positive parameter $\eta$ appears due to the adiabatic application of the external field.
The electric current operators are defined by
        \begin{equation}
        \vj^{\alpha_1 \cdots \alpha_n} =(-1)^n \frac{\partial^n H_\text{BdG}(\bA)}{\partial A^{\alpha_1}\cdots \partial A^{\alpha_n}}\Bigr|_{\bA=0}.\label{generalized_velocity_operator}
        \end{equation}
The first- and second-order ones ($n=1,2$) are called paramagnetic and diamagnetic current density operators, respectively. The detail of the derivation is shown in Appendix~\ref{App_Sec_density_matrix_formulation}.

When the pair potential vanishes, the formula for the NRO conductivity agrees with the prior results for the normal state~\cite{Passos2018,Michishita2020}.
We can check the consistency by replacing $\bA$ with $\bk$ in the derivation.
The replacement is justified by the minimal coupling $\bm{p} \rightarrow \bm{p} -q \bA$ in the normal state, whereas it fails in the superconducting state because particles with opposite charges, that is electron and hole, are treated on equal footing in the BdG formalism.

The failure of the replacement also implies that some useful relations do not hold for the BdG Hamiltonian.
For instance, according to the Hellmann-Feynman theorem ensured in the \textit{normal} state, we obtain the relation between the paramagnetic current operator $J^{\alpha}$ and Berry connection embedded in the band structure as
		\begin{equation}
        J^\alpha_{ab} (\bk) = \frac{\partial}{\partial k_\alpha} E_a \delta_{ab} + i E_{ab} \,\xi^\alpha_{ab}(\bk) , \label{hellmannn_feryman_velocity}
        \end{equation}
in the Bloch representation~\cite{Vanderbilt2018berry}. The Berry connection is defined as
        \begin{equation}
            \xi^\alpha_{ab} (\bk)= i \Braket{u_a (\bk) |\frac{\partial u_b (\bk)}{\partial k_\alpha}}, \label{eq:connection_temp}
        \end{equation}
with the periodic part of the Bloch states $\{ \ket{u_a (\bk)} \}$. This relation is essential for a remarkable simplification of formulas for transport and optical phenomena in the normal state~\cite{Von_Baltz1981,Sipe2000,De_Juan2020,Watanabe2021}. On the other hand, it is unclear whether the nonlinear conductivity of superconductors can be computed similarly to the optoelectronic phenomena in the normal state. 

The failure of the Hellmann-Feynman relation in Eq.~\eqref{hellmannn_feryman_velocity} may make it hard to formulate the optical conductivity in a physically transparent way. The difficulty, however, is eliminated by \textit{vector potential parametrization}. Taking into account the minimal coupling, we introduce the variational parameter $\bm{\lambda}$ by
		\begin{equation}
        \mathcal{H} [\hat{\bm{p}}] \rightarrow \mathcal{H}_\bla = \mathcal{H} [\hat{\bm{p}} - \bm{\lambda} ].
        \end{equation} 
Here and hereafter in this section, we drop ``BdG" of $\mathcal{H}_{\mathrm{BdG}}$ and $H_{\mathrm{BdG}}$ for simplicity.
Since the parameter $\bm{\lambda}$ plays the same role as the (spatially-uniform and time-independent) vector potential, we obtain the relation
		\begin{equation}
            \vj^{\alpha}_{ab} = -\limvec \Braket{a_\bla |\frac{\partial H_\bla}{\partial \lambda_\alpha} | b_\bla},
        \end{equation}
for the paramagnetic current operator. Note that this relation holds even in the BdG formulation. The Hellmann-Feynman relation for the vector potential is obtained as
		\begin{equation}
        \Braket{a_\bla |\frac{\partial H_\bla}{\partial \lambda_\alpha} | b_\lambda} \left[  = - \left( \vj_\bla^\alpha \right)_{ab} \right] = \frac{\partial E_a (\bla)}{\partial \lambda_\alpha} \delta_{ab} - i [\xi^{\lambda_\alpha}, H_\bla]_{ab},  \label{hellmann_feynman_velocity_vector}
        \end{equation}
where we define the connection $\xi^{\lambda_\alpha}$ with the replacement $\partial_{k_\alpha}\to\partial_{\lambda_\alpha}$ in Eq.~\eqref{eq:connection_temp}:
\begin{equation}
    \xi^{\lambda_\alpha}_{ab} = i \Braket{a_\bla |\frac{\partial b_\bla}{\partial \lambda_\alpha}}\label{eq:connection_temp2}.
\end{equation}
The eigenstates $\ket{a_\bla},\ket{b_\bla}$, eigenenergy $E_a (\bla)$, and modified velocity operator $\vj_\bla^\alpha$ are defined on the basis of $H_\bla$.
The quantities defined with the parameter $\bla$ become those with $\bla=\bm{0}$ in the limit $\bla \rightarrow \bm{0}$.
We can apply the Hellmann-Feynman relation to the simplification of the optical conductivity formulas. For instance, the linear optical conductivity and photocurrent response are decomposed into the normal and anomalous contributions with the aid of the $\bla$-parametrization (Appendices~\ref{App_Sec_linear_optical_conductivity} and~\ref{App_Sec_derivation_photocurrent_response}).

Although the introduction of the variational parameter reminds us of the Kohn's classical work~\cite{Resta2018,Kohn1964}, we note that there is a slight difference.
In Ref.~\cite{Kohn1964}, the variational parameter is continuous and can be taken smaller than $2\pi/L$ ($L$ is the linear dimension of the system), whereas we adopt the $2\pi/L$-discretized values for $\bla$ and finally take  the thermodynamic limit $L\to\infty$.
In other words, in a finite system, the variational parameter $\bla$ is defined for the $2\pi/L$-discretized values and the $\bla$ derivative is instead evaluated by the neighboring values for $\bla$.
The finite size effect on Eq.~\eqref{hellmann_feynman_velocity_vector} vanishes after the thermodynamic limit.
Note that all physical observables have $2\pi/L$-periodic $\bla$ dependence in the normal state, while generally they do not in the superconducting state based on the BdG formalism with a fixed pair potential.
The reason is the following: In the normal state, the introduction of $\bla$ just shifts the wave number, and $\bla$ dependence of physical quantities vanishes after the $\bm{k}$ summation.
By contrast, in the superconducting state, $\bla$ corresponds to the center-of-mass momentum of the Cooper pairs and, thus, the states with different $\bla$ should be distinguished from each other.
While the value of $\bla$ minimizing the free energy (denoted by $\bla=\bm{0}$ in this paper) is realized in equilibrium, the derivatives of $\bla$ have physical implications such as the superfluid weight as we see below.
Thus, the vector potential parametrization offers us a good means to extract physical properties intrinsic to superconductors.

\subsection{Anomalous photocurrent responses} 
\label{Secsub_anomalous_photocurrent_response}

Emergence of the superconductivity leads to unique optical phenomena which we call anomalous optical responses. For example, we investigate the photocurrent response given by the NRO conductivity $\sigma^{\alpha;\beta\gamma} (0;\Omega,-\Omega)$, and frequency dependence will be suppressed in this section.

The real and imaginary parts denote the photocurrent conductivity induced by linearly-polarized and circularly-polarized-lights, respectively~\cite{Sturman1992Book}.
According to the prior studies investigating the normal state, the mechanism for the photocurrent generation and its dependence on the polarization state of light are closely related to the temporal symmetry such as \T{} and \PT{} symmetries~\cite{Von_Baltz1981,Sipe2000,Zhang2019,De_Juan2020,Holder2020,Watanabe2021}.
The same classification applies to the superconducting state as revealed in our derivation.

Leaving the derivation to Appendix~\ref{App_Sec_derivation_photocurrent_response}, we show the formulas for the NRO response used in the following calculation.
In the gapful superconductors at low temperatures, the total photocurrent conductivity is given by the two components; 
		\begin{equation}
            \sigma = \sigma_\text{n} + \sigma_\text{a}. \label{total_photocurrent_conductivity_gapped}
        \end{equation}
The former ($\sigma_\text{n}$) is a normal photocurrent which is a counterpart of the known photocurrent in the normal state, while the latter new contribution ($\sigma_\text{a}$) represents an \textit{anomalous photocurrent} unique to the superconducting state.

The normal part consists of the four contributions
		\begin{equation}
        \sigma_\text{n} = \sigma_\text{Einj} + \sigma_\text{Minj}+ \sigma_\text{shift} + \sigma_\text{gyro},
        \end{equation}
which are termed electric injection current, magnetic injection current, shift current, and gyration current, respectively. The formulas read
        \begin{align}
        &\sigma^{\alpha;\beta\gamma}_\text{Einj}=   -\frac{i\pi}{8\eta}  \sum_{a\neq b} \left( \vj^{\alpha}_{aa}-\vj^{\alpha}_{bb}  \right) \Omega_{ba}^{\lambda_\beta \lambda_\gamma} F_{ab}, \label{electric_injection} \\
        &\sigma^{\alpha;\beta\gamma}_\text{Minj}=   \frac{\pi}{4\eta}  \sum_{a \neq b} \left( \vj^{\alpha}_{aa}-\vj^{\alpha}_{bb}  \right) g_{ba}^{\lambda_\beta\lambda_\gamma} F_{ab},\label{magnetic_injection} \\
        &\sigma^{\alpha;\beta\gamma}_\text{shift} = -\frac{\pi}{4}   \sum_{a\neq b}  \Im{\Biggl[ [D_{\lambda_\alpha} \xi^{\lambda_\beta}]_{ab}\xi^{\lambda_\gamma}_{ba} + [D_{\lambda_\alpha} \xi^{\lambda_\gamma}]_{ab}\xi^{\lambda_\beta}_{ba}  \Biggr]} F_{ab},\label{shift_current_berry_connection}\\
        &\sigma^{\alpha;\beta\gamma}_\text{gyro} = -\frac{i\pi}{4}   \sum_{a\neq b}  \Re{\Biggl[ [D_{\lambda_\alpha} \xi^{\lambda_\beta}]_{ab}\xi^{\lambda_\gamma}_{ba} - [D_{\lambda_\alpha} \xi^{\lambda_\gamma}]_{ab}\xi^{\lambda_\beta}_{ba}  \Biggr]} F_{ab}.\label{gyration_current_berry_connection}
        \end{align}
$F_{ab} = f_{ab}\, \delta (\Omega - E_{ba})$ means the Pauli exclusion principle at the optical transition.
We also defined the geometric quantities such as Berry curvature ($\Omega_{ab}^{\lambda_\alpha\lambda_\beta}$), quantum metric ($g_{ab}^{\lambda_\alpha\lambda_\beta}$), and the covariant derivative ($D_{\lambda_\alpha}$). The covariant derivative is associated with the equi-energy Berry connection where bra- and ket-states satisfy $E_a = E_b$.
Note that we finally take the limit $\bla \rightarrow \bm{0}$.
The shift current formula is in agreement with that derived in Ref.~\cite{Xu2019}.
We also note that the known formulas for the normal photocurrent response~\cite{Watanabe2021} are reproduced by the replacement of $\bla$ with the crystal momentum $\bk$.
The replacement is justified only in the normal state where the pair potential $\Delta$ is zero.

Because of the inequivalence of $\bla$ and $\bk$, the anomalous NRO conductivity appears in the superconducting state.
The anomalous part consists of two contributions
		\begin{equation}
            \sigma_\text{a} = \sigma_\text{NRSF} +  \sigma_\text{CD}.\label{anomalous_photocurrent_total}
        \end{equation}
The expressions are given by 
        \begin{align}
            &\sigma^{\alpha;\beta\gamma}_\text{NRSF} 
            = \limvec -\frac{1}{2\Omega^2}  \partial_{\lambda_\alpha}  \partial_{\lambda_\beta} \partial_{\lambda_\gamma} F_{\bm \lambda}, \label{anomalous_photocurrent_zeroth}\\
            &\sigma^{\alpha;\beta\gamma}_\text{CD} = \limvec \frac{1}{4\Omega^2} \partial_{\lambda_\alpha} \left[  \sum_{a\neq b }  \vj^\beta_{ab}\vj^{\gamma}_{ba} f_{ab}  \left( \frac{1}{\Omega -E_{ab}} + \frac{1}{E_{ab}}  \right)     \right]. \label{anomalous_photocurrent_linear}
        \end{align}
In Eq.~\eqref{anomalous_photocurrent_zeroth}, $F_{\bm{\lambda}}$ is the free energy of the BdG Hamiltonian.
The second-order derivative of the free energy is the superfluid density $\rho_\text{s}^{\alpha\beta} = \limvec \partial_{\lambda_\alpha}  \partial_{\lambda_\beta}  F_{\bm{\lambda}}$, which supports the macroscopic quantum coherence of the superconducting state~\cite{Tinkham2004introduction}.
Thus, $\sigma^{\alpha;\beta\gamma}_\text{NRSF}$ is determined by the nonreciprocal correction to the superfluid density $\rho_\text{s}^{\alpha\beta}$, which we call \textit{nonreciprocal superfluid density} $ f^{\alpha\beta\gamma} = \limvec \partial_{\lambda_\alpha}  \partial_{\lambda_\beta}  \partial_{\lambda_\gamma} F_{\bm{\lambda}}$. The anomalous term in Eq.~\eqref{anomalous_photocurrent_zeroth} is rewritten by
		\begin{equation}
           \sigma^{\alpha;\beta\gamma}_\text{NRSF} = -\frac{1}{2\Omega^2} f^{\alpha\beta\gamma}.\label{nonreciprocal_superfluid_density_effect}
        \end{equation}
The nonreciprocal superfluid density $\hat{f}$ is a totally-symmetric rank-3 tensor.
It shows the odd-parity under the \Pa{} and \T{} symmetry operations, whereas the even-parity under the \PT{} operation.

Another contribution in Eq.~\eqref{anomalous_photocurrent_linear} is named \textit{conductivity derivative effect}, because the formula roughly corresponds to the $\bla$-derivative of the linear optical conductivity.
Supporting this argument, the formula in the low-frequency limit ($\Omega \rightarrow 0$) leads to the $\bla$-derivative of the total Berry curvature $ \sum_a \epsilon_{\beta\gamma\delta} \Omega^{\lambda_\delta}_{a} f_a $, which gives the anomalous Hall conductivity~\cite{Niu1984}.
Thus, we can recast Eq.~\eqref{anomalous_photocurrent_linear} as
		\begin{align}
           \sigma^{\alpha;\beta\gamma}_\text{CD} \rightarrow \sigma^{\alpha;\beta\gamma}_\text{sCD} 
           &=  \limvec  \frac{i}{4\Omega} \epsilon_{\beta\gamma\delta} \partial_{\lambda_\alpha} \left( \sum_a \Omega^{\lambda_\delta}_{a} f_a  \right),\\
           &\equiv \frac{i}{4\Omega} \epsilon_{\beta\gamma\delta} B_{d}^{\,\alpha \delta},\label{Berry_curvature_derivative_effect}
        \end{align}
in the low-frequency regime. Here, we represented the conductivity derivative effect in the static limit ($\sigma^{\alpha;\beta\gamma}_\text{sCD}$) by \textit{Berry curvature derivative} $\hat{B}_d$ and suppressed $O(\Omega^0)$ terms.
The temporal symmetry of $\hat{B}_d$ is contrasting with the nonreciprocal superfluid density $\hat{f}$; $\hat{B}_d$ is \T{}-even and \PT{}-odd.

The anomalous photocurrent responses are particularly of interest due to the giant photo-electric conversion for the low-frequency light.
The anomalous photocurrent response shows high performance for the light with frequency up to the energy scale of the superconductivity $(\Omega \lesssim \text{0.1\,-\,1 THz}$).
The \PT{} symmetric parity-breaking superconductors show the divergent response as large as $O(\Omega^{-2})$ due to the nonreciprocal superfluid density, whereas the \T{} symmetric superconductors host the $O(\Omega^{-1})$-diverging response arising from the Berry curvature derivative.
Furthermore, the photocurrent converted from the low-frequency light may be carried by Cooper pairs and hence flow without undesirable Joule heating.
The energy-saving property originates from the quantum mechanical nature of the superconductors and cannot be achieved by existing photo-electric converters such as the p-n junction of semiconductors.

\subsection{Anomalous contributions to general NRO responses} 
\label{Secsub_Green_function_results}

In Sec.~\ref{Secsub_anomalous_photocurrent_response}, the anomalous and normal photocurrent responses are detailed on the basis of the gapped parity-breaking superconducting state.
The low-frequency divergent behavior is similarly observed in other NRO responses such as the second harmonic generation.
In this subsection, the anomalous contribution to the NRO responses is obtained by the Green function method. Although the derivation is left to Appendix~\ref{App_Sec_Green_function_method_derivation}, the low-frequency anomalous contribution reads
		\begin{align}
            &\sigma^{\alpha ;\beta \gamma} \left(\omega_1+\omega_2 ; \omega_{1}, \omega_{2}\right) \notag \\
            &=  \sigma^{\alpha ;\beta \gamma}_\text{NRSF} \left(\omega_1+\omega_2 ; \omega_{1}, \omega_{2}\right) + \sigma^{\alpha ;\beta \gamma}_\text{sCD} \left(\omega_1+\omega_2 ; \omega_{1}, \omega_{2}\right),\label{general_anomalous_divergent_NRO_conductivity}
        \end{align}
where we suppressed $O(\omega_1^a \omega_2^b)$ terms ($a+b \geq 0$).
The formula is given by the nonreciprocal superfluid density and the static conductivity derivative effects
        \begin{align}
        &\sigma^{\alpha ;\beta \gamma}_\text{NRSF} =  \frac{1}{2\omega_1 \omega_2} f^{\alpha\beta\gamma},\label{NRSF_effect_green_function}\\
        &\sigma^{\alpha ;\beta \gamma}_\text{sCD} =  -\frac{i}{4} \limvec \left( \frac{1}{\omega_2} \partial_{\lambda_\beta} \sigma_{\alpha\gamma}^{(\bla)} + \frac{1}{\omega_1}  \partial_{\lambda_\gamma} \sigma_{\alpha\beta}^{(\bla)} \right).
        \end{align}
The tensor $\hat{\sigma}^{(\bla)}$ represents the regular linear conductivity.
Since we formulated the NRO conductivity by the Green function method with neglecting vertex corrections, the formulas also hold in the presence of a self-energy correction such as that originates from disorder scattering.
The regular linear conductivity consists of the Drude and Berry curvature contributions.
Thus, when the conductivity tensor $\sigma_{\alpha\beta}^{(\bla)}$ is decomposed into the symmetric and antisymmetric components under the permutation $\alpha \leftrightarrow \beta$, the static conductivity derivative is further decomposed into the Drude derivative $\hat{D}_d$ and Berry curvature derivative $\hat{B}_d$
		\begin{align}
        &\partial_{\lambda_\gamma} \sigma_{\alpha\beta}^{(\bla)} \notag \\ 
            &=\partial_{\lambda_\gamma} \left[ \frac{1}{2} \left( \sigma_{\alpha\beta}^{(\bla)} + \sigma_{\beta\alpha}^{(\bla)}  \right) + \frac{1}{2} \left( \sigma_{\alpha\beta}^{(\bla)} - \sigma_{\beta\alpha}^{(\bla)}  \right)  \right],\\
            &= D_d^{\gamma;\alpha\beta} + \epsilon_{\alpha\beta\delta}  B_d^{\gamma\delta}.
        \end{align}
The Drude derivative defined by Eq.~\eqref{App_cond_deriv_PT_sym} vanishes without quasiparticle excitation.
In contrast to the Berry curvature derivative (\T{}-even and \PT{}-odd), the Drude derivative is forbidden by the \T{} symmetry but allowed in the \PT{} symmetric systems.
As a result, the static conductivity derivative effect is rewritten by
\begin{align}
    &\sigma^{\alpha ;\beta \gamma}_\text{sCD}\notag \\
    &= -\frac{i}{4}  \left[ \frac{1}{\omega_2} \left( D_d^{\beta;\alpha\gamma} + \epsilon_{\alpha\gamma\delta}  B_d^{\beta\delta}  \right) + \frac{1}{\omega_1} \left( D_d^{\gamma;\alpha\beta} + \epsilon_{\alpha\beta\delta}  B_d^{\gamma\delta} \right) \right]. \label{general_static_conductivity_derivative}
    \end{align}

The obtained expressions reveal that the nonreciprocal superfluid density and Berry curvature derivative contribute to the general NRO conductivity as well as the photocurrent generation.
Furthermore, the equation~\eqref{general_anomalous_divergent_NRO_conductivity} includes the Fermi surface effect such as the Drude derivative contribution, while the gapped superconductor is assumed in Sec.~\ref{Secsub_anomalous_photocurrent_response}.
If the Fermi-surface contribution is neglected, we obtain the relation of the Berry curvature derivative 
		\begin{equation}
        \epsilon_{\beta\gamma\delta} B_d^{\alpha\delta} = \epsilon_{\alpha\gamma\delta} B_d^{\beta\delta} - \epsilon_{\alpha\beta\delta} B_d^{\gamma\delta}. 
        \end{equation}
Taking $\omega_1=-\omega_2 = \Omega$ in Eq.~\eqref{general_anomalous_divergent_NRO_conductivity} and assuming the gapped superconducting state, we reproduce the Berry curvature derivative effect in Eq.~\eqref{Berry_curvature_derivative_effect}.

We comment on the tensor symmetry of the above-mentioned quantities induced by the parity-breaking superconductivity.
First, the nonreciprocal superfluid density $\hat{f}$ is a totally-symmetric rank-3 polar tensor and thus allowed in the noncentrosymmetric crystal point groups except for the several high-symmetric chiral groups where $C_n\,(n\geq 4)$ rotation symmetry holds [$432\, (O)$, $422$\, ($D_{4}$), $622$\, $(D_6)$].
Next, we consider the static conductivity derivative. The Berry curvature derivative $\hat{B}_d$ is the axial rank-2 tensor and hence allowed in the gyrotropic groups~\cite{Halasyamani1998noncentrosymmetric}.
Here, we denote the gyrotropic groups as the noncentrosymmetric crystal point groups other than the three crystal point groups $\bar{6}\, (C_{3h}),~\bar{6}m2\, (D_{3h}),~\bar{4}3m\, (T_{d})$.
The Drude derivative $\hat{D}_d$ is the direct product of the rank-1 and rank-2 polar tensors.
Thus, it has the same symmetry as that of the piezoelectric response coefficient and allowed in all the noncentrosymmetric crystal point groups except for the cubic and chiral group $432\,(O)$.

Finally,
we provide a classification table of photocurrent generation (Table~\ref{Table_photocurrent_classification}) to summarize the results obtained by the density matrix (Sec.~\ref{Secsub_anomalous_photocurrent_response}) and Green function (Sec.~\ref{Secsub_Green_function_results}) methods. Mechanism of normal and anomalous photocurrent is classified based on the temporal symmetry of the system and the polarization state of light. 

    \begin{table*}[htbp]
        \caption{
        Classification of normal and anomalous photocurrent responses.
        Parities under the \T{} and \PT{} symmetry operations are denoted by $\pm$.
        In the column `$\bm{e}$' we specify the linearly-polarized-light-induced (\LP) or circularly-polarized-light-induced (\CP) photocurrent. The check mark (\cm) in the column `FS (Fermi Surface effect)' indicates the photocurrent which vanishes in the absence of the Fermi surface.
        As for the normal photocurrent, we reproduce the classification performed in Ref.~\cite{Watanabe2021} where the normal state is assumed. On the other hand, the anomalous photocurrent is unique to superconductors.}
    \label{Table_photocurrent_classification}
    \centering
    \vspace{5pt}
    \begin{tabular}{lccccc}\hline \hline
    Mechanism&\T{}&\PT{}&$\bm{e}$&FS&Ref.\\ \hline
    (Normal photocurrent)&&&&&\\
    Drude                               &  - & + & \LP & \cm &\cite{Holder2020} \\
    Berry curvature dipole              &  + & - & \CP & \cm &\cite{Moore2010} \\
    Intrinsic Fermi surface (electric)  &  + & - & \CP & \cm &\cite{De_Juan2020} \\
    Intrinsic Fermi surface (magnetic)  &  - & + & \LP & \cm &\cite{Watanabe2021} \\
    Injection current (electric)       &  + & - & \CP &     &\cite{Sipe2000} \\
    Injection current (magnetic)        &  - & + & \LP &     &\cite{Zhang2019} \\
    Shift current                       &  + & - & \LP &     &\cite{Von_Baltz1981,Sipe2000} \\
    Gyration current                    &  - & + & \CP &     &\cite{Watanabe2021,Ahn2020} \\ 
    \\ 
    (Anomalous photocurrent)&&&&&\\
    Nonreciprocal superfluid density         &  - & + & \LP &     &\textbf{This work} \\
    Berry curvature derivative          &  + & - & \CP &     &\textbf{This work} \\
    Drude derivative          &  - & + & \CP & \cm  &\textbf{This work} \\\hline \hline
    \end{tabular}
    \end{table*}

\section{Generalization to many-body systems}
\label{Sec_manybody}

In Sec.~\ref{Sec_formulation}, we derived the anomalous NRO responses from the second-order perturbation theory.
In this section, we derive a more general formula for anomalous responses unique to superconductors, which applies to interacting systems.

We consider the many-body Hamiltonian comprising parameters $\bm{\lambda}$ which adiabatically evolve in time.
We assume that the ground state is separated from the excited states by a finite energy gap.
We also assume the zero temperature.
The secular equation of the Hamitonian $H= H (\bla(t))$ reads
        \begin{equation}
            H (\bla (t)) \ket{a_\bla} = \varepsilon_a (\bla) \ket{a_\bla},
        \end{equation}
where the instantaneous eigenstates and eigenvalues are introduced.
If the many-body wavefunction is initially in the ground state $\ket{0}$ and adiabatically evolves in time, the wavefunction is obtained as~\cite{Thouless1983}
        \begin{align}
            \ket{\psi (t)} 
                &= \ket{0_\bla} + \sum_{a\neq 0} \frac{-i}{\varepsilon_0 (\bla) - \varepsilon_a (\bla) } \ket{a_\bla} \Braket{a_\bla | \partial_t 0_\bla},\\
                &= \ket{0_\bla} - \sum_{a\neq 0} \frac{1}{\varepsilon_0 (\bla) - \varepsilon_a (\bla) } \ket{a_\bla} \Xi_{a0}^{\lambda_\alpha} \dot{\lambda_\alpha}.
        \end{align}
with the many-body Berry connection $\Xi_{a0}^{\lambda_\alpha}\equiv i\braket{a_{\bm{\lambda}}|\partial_{\lambda_\alpha}0_{\bm{\lambda}}}$.
The operator $\partial_{\lambda_\alpha} H (\bla(t))$ is evaluated by the perturbed wavefunction $\ket{\psi(t)}$ as
        \begin{align}
            \Braket{\frac{\partial H (\bla (t))}{\partial \lambda_\alpha}} 
            &=\Braket{\psi(t) | \frac{\partial H (\bla (t))}{\partial \lambda_\alpha} | \psi(t)},\\
            &= \frac{\partial \varepsilon_0 (\bla)}{\partial \lambda_\alpha} + \Omega^{\lambda_\alpha\lambda_\beta} \dot{\lambda_\beta}.
        \end{align}
The many-body Berry curvature is
        \begin{align}
            \Omega^{\lambda_\alpha\lambda_\beta}
                &= 2 \Im{\,[ \Braket{\partial_{\lambda_\alpha} 0_\bla | \partial_{\lambda_\beta} 0_\bla}]},\\
                &= \sum_{a \neq 0} i \left(  \Xi^{\lambda_\alpha}_{0a} \Xi^{\lambda_\beta}_{a0} -\Xi^{\lambda_\alpha}_{a0} \Xi^{\lambda_\beta}_{0a}  \right).
        \end{align}
When we take the vector potential $\bA$ as the adiabatic parameter, the expectation value of the electric current is obtained as
        \begin{align}
            \Braket{\mathcal{J}^\alpha (t)} 
                &= -\Braket{\frac{\partial H (\bA (t))}{\partial A_\alpha}},\\
                &= - \frac{\partial \varepsilon_0 (\bA)}{\partial A_\alpha} - \Omega^{A_\alpha A_\beta} \dot{A_\beta}.\label{generalized_anomalous_response_adiabatic}
        \end{align}
The first term is related to the generalized Drude weight evaluated at the zero temperature~\cite{Drudeweight}.
When the ground state energy $\varepsilon_0 (\bA)$ is expanded by the vector potential, we obtain
        \begin{align}
            \varepsilon_0 (\bA) 
                &=\varepsilon_0 (\bm{0}) + \frac{1}{2}\left. \frac{\partial^2 \varepsilon_0 (\bA) }{\partial A_\alpha \partial A_\beta}\right|_{ \bA = \bm{0}} A_\alpha A_\beta \notag \\
                &+ \frac{1}{6}\left. \frac{\partial^3 \varepsilon_0 (\bA) }{\partial A_\alpha \partial A_\beta \partial A_\gamma}\right|_{ \bA = \bm{0}} A_\alpha A_\beta A_\gamma+ O(\bA^4).\label{generalized_anomalous_response_adiabatic_NRSF}
        \end{align}
The first-order term is dropped because the electric current is absent in the unperturbed ground state.
$\partial_{A}^2 \varepsilon_0 (\bA)$ and $\partial_{A}^3 \varepsilon_0 (\bA)$ are derivatives of the free energy in the thermodynamic limit which means linear and nonreciprocal superfluidity.
The agreement between the Drude weight and superfluid density (Meissner weight) is a property of the gapful superconductor at the zero temperature~\cite{Scalapino1993}.
Converting the vector potential into the electric field $\bm{E} = -\dot{\bA}$, we obtain the anomalous linear and nonreciprocal optical responses in the many-body representation. For instance, the response formula for the nonreciprocal part is written by
    \begin{align}
        \sigma_\text{NRSF}^{\alpha;\beta\gamma} (\omega;\omega_1,\omega_2)
            &=  \frac{1}{\omega_1  \omega_2 } \left.\frac{\partial^3 \varepsilon_0 (\bA) }{\partial A_\alpha \partial A_\beta \partial A_\gamma}\right|_{ \bA = \bm{0}},\label{generalized_anomalous_response_adiabatic_NRSF}
    \end{align}
where $\omega = \omega_1 +\omega_2$ and the adiabaticity parameter $\eta$ is suppressed.
When we apply the formula to one-body Hamiltonian, Eqs.~\eqref{NRSF_effect_green_function} and~\eqref{App_superfluid_density} are reproduced.

Similarly, we formulate the Berry curvature derivative effect on the basis of the many-body Hamilonian.
The many-body Berry curvature $\Omega^{A_\alpha A_\beta}$ is parametrized by the vector potential, and the Berry curvature derivative appears in the first-order expansion coefficient.
        \begin{equation}
            \Omega^{A_\alpha A_\beta} = \left.\Omega^{A_\alpha A_\beta}\right|_{\bA = \bm{0}} +\left.\frac{\partial \Omega^{A_\alpha A_\beta } }{\partial A_\gamma}\right|_{\bA = \bm{0}} A_\gamma + O(\bA^2).
        \end{equation}
When a one-body Hamiltonian is assumed, the many-body Berry curvature leads to the one-body expression.
Accordingly, the first-order expansion coefficient $\partial_{A_\gamma} \Omega^{A_\alpha A_\beta}$ reproduces the Berry curvature derivative.
The Berry curvature derivative contributes to the nonreciprocal conductivitity as~\cite{noteresta}.
        \begin{align}
         \sigma_\text{sCD}^{\alpha;\beta\gamma} (\omega;\omega_1,\omega_2)
            &=  \frac{-i}{2\omega_2} \left. \frac{\partial \Omega^{A_\alpha A_\beta } }{\partial A_\gamma}\right|_{\bA = \bm{0}}+  \frac{-i}{2\omega_1}\left. \frac{\partial \Omega^{A_\alpha A_\gamma } }{\partial A_\beta}\right|_{\bA = \bm{0}} .\label{generalized_anomalous_response_adiabatic_BCD}
        \end{align}
Thus, the obtained formulas~\eqref{generalized_anomalous_response_adiabatic_NRSF} and \eqref{generalized_anomalous_response_adiabatic_BCD} are consistent with the low-frequency limit of the anomalous NRO response formulas shown in the previous section.

Although we here consider the vector potential effect on the electric current up to the second-order, higher-order anomalous contributions are similarly obtained as $O(|\bA|^n)$ or $O(|\bA|^{n-1} |\dot{\bA}|)$ terms ($n\geq 3$), which are higher-order $\bA$-derivative of the superfluid density and Berry curvature, respectively.
These contributions may dominate the nonlinear optical responses such as the high-order harmonic generation in the low-frequency regime, while the resonant contributions have been intensively studied~\cite{Shimano2020}.

Finally, we give a few comments. We generalized the formulas of the anomalous NRO responses to those for the many-body Hamiltonian [Eqs.~\eqref{generalized_anomalous_response_adiabatic_NRSF},\, \eqref{generalized_anomalous_response_adiabatic_BCD}].
The nonreciprocal contributions in Eq.~\eqref{generalized_anomalous_response_adiabatic} can cover a broad range of electromagnetic responses, while only the photo-electric field is treated in Sec.~\ref{Sec_formulation}.
For instance, with a spatially-nonuniform vector potential, the anomalous nonreciprocal contribution leads to the nonreciprocal Meissner response~\cite{HDY_meissner}.

\section{Model study of anomalous NRO responses} 
\label{Sec_model_study}

Here, we demonstrate the anomalous NRO responses with numerical calculations.
In this section, we consider the parity-breaking superconductor classified into Case (III) (see Sec.~\ref{Sec_symmetry_analysis}), where superconductivity itself breaks the \Pa{} symmetry, by referring to the recent proposal of the spontaneous parity-mixed superconductivity in a heavy fermion superconductor UTe$_2$~\cite{Ishizuka2021}.

With the model Hamiltonian introduced in Sec.~\ref{Secsub_model_hamiltonian}, we investigate the normal and anomalous contributions to the photocurrent response and corroborate our classification based on the preserved temporal symmetry (Sec.~\ref{Secsub_anomalous_photocurrent_model_study}).
The dominant role of the anomalous contribution appears in the numerical calculations of the photocurrent and second harmonic generations (Sec.~\ref{Secsub_anomalous_PGE_SHG_model_study}).
Finally, we investigate scattering rate dependence of the anomalous NRO response (Sec.~\ref{Secsub_scattering_rate}).

\subsection{Model Hamiltonian}
\label{Secsub_model_hamiltonian} 

The model Hamiltonian consists of spinful fermions on the two-dimensional square lattice.
In the tight-binding approximation, the normal part $H_\text{N}$ of $H_\text{BdG}$ reads
        \begin{equation}
            H^{\bk}_\text{N} = \left( \varepsilon_0- \mu  \right) + V \rho_x + \bm{g}\cdot \bm{\sigma} \rho_z,
        \end{equation}
at each crystal momentum $\bk$.
Pauli matrices $\bm{\sigma}$ and $\bm{\rho}$ respectively denote the spin and crystal sublattice degrees of freedom.
The model is constructed on the square network where two crystal sublattices are placed in the checkerboard pattern.
The Hamiltonian comprises the chemical potential $\mu$ and the intra- and intersublattice hoppings defined as 
		\begin{align}
        &\varepsilon_0 = -4 t_1 \cos{k_x} \cos{k_y},\\
        &V = -2 t_2 \left( \cos{k_x}+  \cos{k_y} \right).
        \end{align}
Assuming the locally noncentrosymmetric structure, the staggered spin-orbit coupling is given by the local Rashba type~\cite{KaneMele2005,Yanase2014,Zelezny2014}
		\begin{equation}
        \bm{g} = 4\alpha \left( \cos{k_x}\sin{k_y},-\sin{k_x}\cos{k_y},0  \right).
        \end{equation}

The parity-mixed pair potential consists of the $s$-wave component $\psi_\bk = \Delta_\text{e}$ and $p$-wave component $\bm{d}_\bk = 4\Delta_\text{o} \sin{k_x}\cos{k_y} \hat{z}$.
For a \T{} symmetric parity-breaking superconductor, the pair potential is taken in the $s+p$-wave form
		\begin{equation}        
        \Delta_\bk i\sigma_y = \psi_\bk + \bm{d}_\bk\cdot \bm{\sigma},
        \end{equation}
at each crystal momentum $\bk$, while the $s+ip$-wave type pair potential is given by
		\begin{equation}
            \Delta_\bk i\sigma_y = \psi_\bk + i \bm{d}_\bk\cdot \bm{\sigma},       
        \end{equation}
for a \PT{} symmetric superconductor.
The superconducting state is the admixture of the $A_g$- and $E_u$-type pairings in terms of the irreducible representation of the point group $D_{4h}$ ($4/mmm$).
The preserved unitary symmetry operations in the parity-breaking superconducting state are characterized by the $C_{2v}$ ($m2m$) symmetry, whose two-fold rotation axis is the $y$-axis.
This symmetry allows the NRO conductivity, $\sigma^{y;yy}$, $\sigma^{y;xx}$, $\sigma^{x;xy}$, $\sigma^{x;yx}$, $\sigma^{z;yz}$, $\sigma^{z;zy}$, and $\sigma^{y;zz}$.
The last three components having the index $z$ are not considered because the two-dimensional model is adopted.

The parameters are $\mu=-4$, $\alpha= 0.3$, $t_1 = 0.6$, $t_2 = 1.0$, $\Delta_\text{e} + 4\Delta_\text{o} =  0.1$, and ${\Delta_\text{e}}/{\Delta_\text{o}} =  4$.
For a quantitative estimate, we set $t_2= \mr{1}{eV}$ and calculate the response coefficients in the SI unit.
Thus, the numerical results of the NRO conductivity are given in the unit A$\cdot $V$^{-2}$.
Although the model Hamiltonian is two-dimensional, we convert the obtained NRO conductivity into the three-dimensional value by using the thickness of the two-dimensional net $l = \mr{1}{nm}$.
The numerical calculations are performed on the $N^2$-discretized Brillouin zone.
For numerical convergence, we introduce a phenomenological scattering rate $\gamma$ (Appendix~\ref{App_Sec_Phenomenology}) and a finite temperature $T$ for the Fermi-Dirac distribution function.

\subsection{Normal and anomalous photocurrent responses}
\label{Secsub_anomalous_photocurrent_model_study}

\begin{figure*}[htbp]
    \centering
    \includegraphics[width=0.80\linewidth,clip]{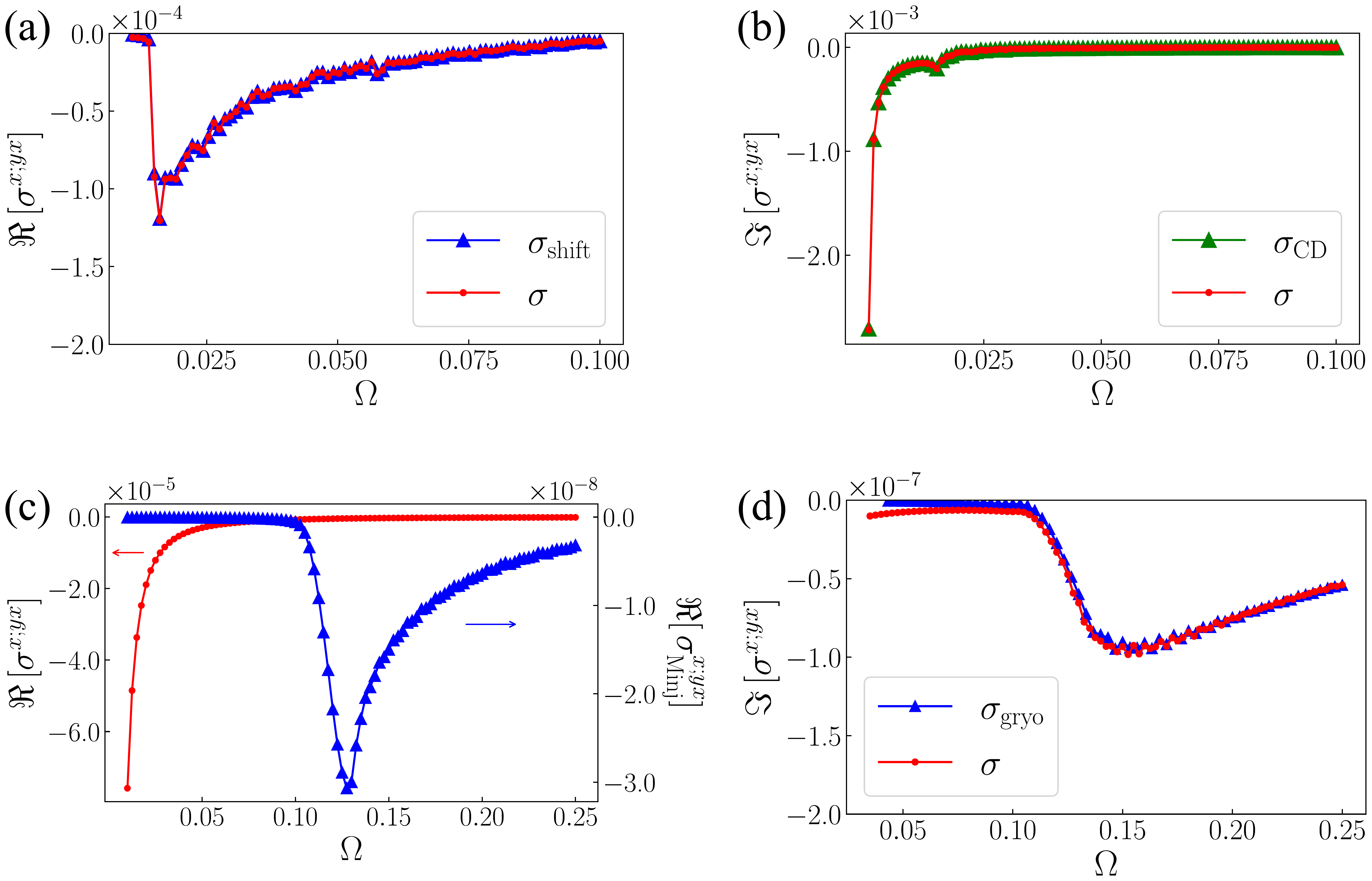}
    \caption{
        Frequency dependence of the photocurrent conductivity $\sigma^{x;xy}$
        in (a,b) a \T{} symmetric parity-breaking superconductor and  (c,\,d) a \PT{} symmetric parity-breaking superconductor.
        (a) and (c) plot the real part while (b) and (d) are the imaginary part.
        The total photocurrent conductivity ($\sigma$) is shown in red, while the normal and anomalous photocurrent conductivities are shown in blue and green, respectively.
        The calculation is performed with $N=4000$ and $T=10^{-4}$. We also adopted $\gamma = 10^{-4}$ for the \T{} symmetric case and $\gamma = 3\times 10^{-3}$ for the \PT{} symmetric case.
        }
\label{Fig_photocurrent_comparison_xyx}
\end{figure*}

Here, numerical analysis of the photocurrent conductivity $\sigma^{x;yx} (0;\Omega,-\Omega)$ is presented.
The frequency dependence of the conductivity is implicit unless otherwise mentioned.
Figure~\ref{Fig_photocurrent_comparison_xyx} plots the frequency dependence of the photocurrent conductivity $\sigma^{x;yx}$ under the linearly-polarized and circularly-polarized lights. To distinguish the normal and anomalous photocurrent responses, we calculate each contribution as well as the total photocurrent conductivity.

In the case of the \T{} symmetric system, the real part $\Re{\,[\sigma^{x;yx}]}$ for the linearly-polarized light in Fig.~\ref{Fig_photocurrent_comparison_xyx}(a) stems from the normal photocurrent mechanism, that is, the shift current $(\sigma_\text{shift})$~\cite{Von_Baltz1981,Sipe2000}.
Since the normal photocurrent requires quasiparticle excitations, the contribution vanishes in the low-frequency regime $(\Omega \lesssim 0.01)$ due to the superconducting gap. 
On the other hand, the imaginary part $\Im{\,[\sigma^{x;yx}]}$ for the circularly-polarized light is solely determined by the anomalous contribution, namely the conductivity derivative $(\sigma_\text{CD})$ [Fig.~\ref{Fig_photocurrent_comparison_xyx}(b)].
In agreement with Eq.~\eqref{Berry_curvature_derivative_effect}, we clearly observe the low-frequency divergence proportional to $\Omega^{-1}$ in the circular photocurrent conductivity $\Im{\,[\sigma^{x;yx}]}$.
Although the normal photocurrent called the electric injection current $(\sigma_\text{Einj})$ also contributes to the circular photocurrent conductivity, it gives a negligible contribution compared to the anomalous contribution.
This result implies that the anomalous mechanism provides distinctive optoelectronic properties of parity-breaking superconductors.

Similarly, it is shown that the anomalous photocurrent gives rise to the dominant photocurrent creation in the \PT{} symmetric parity-breaking superconductors [Fig.~\ref{Fig_photocurrent_comparison_xyx}(c,\,d)].
Owing to the \PT{} symmetry, the situation is in contrast to the \T{} symmetric superconductors. The circular photocurrent conductivity $\Im{\,[\sigma^{x;yx}]}$ is determined by the normal photocurrent called gyration current $(\sigma_\text{gyro})$~\cite{Ahn2020,Watanabe2021} [Fig.~\ref{Fig_photocurrent_comparison_xyx}(d)], while the linear photocurrent $\Re{\,[\sigma^{x;yx}]}$ is given by the anomalous contribution, that is due to the nonreciprocal superfluid density effect $(\sigma_\text{NRSF})$, with a negligible normal photocurrent originating from the magnetic injection current $(\sigma_\text{Minj})$~\cite{Zhang2019} [Fig.~\ref{Fig_photocurrent_comparison_xyx}(c)].
The low-frequency divergence of $\Re{\,[\sigma^{x;yx}]}$ is inversely proportional to $\Omega^{2}$ in agreement with the formula~\eqref{nonreciprocal_superfluid_density_effect}.

Here, we compare the photocurrent conductivity and the numerically calculated nonreciprocal superfluid density to make the analytical result in Eq.~\eqref{nonreciprocal_superfluid_density_effect} more compelling.
Figure~\ref{Fig_NRSF_PGE_comparison} shows the comparison between $\Re{[\,\Omega^2\sigma^{x;yx}]} $ and the nonreciprocal superfluid density $f^{xyx}$.
In agreement with the analytical result, the NRO conductivity multiplied by the square of frequency $\Omega^2$ asymptotically approaches to the nonreciprocal superfluid density in the low-frequency limit.
Since $\Re{[\,\sigma^{x;yx}]}$ includes a non-divergent part of the conductivity derivative effect in Eq.~\eqref{anomalous_photocurrent_linear}, the dispersion in $\Re{[\,\Omega^2 \sigma^{x;yx}]}  \propto \Omega^2 $ is observed.
The agreement between the analytical and numerical results indicates that the nonreciprocal superfluid density is an indicator of the low-frequency NRO response. Conversely, the NRO measurements such as the second harmonic generation and photocurrent response may provide a quantitative estimate of the nonreciprocal superfluid density.

\expandafter\ifx\csname iffigure\endcsname\relax
        \begin{figure}[htbp]
        \centering
        \includegraphics[width=0.90\linewidth,clip]{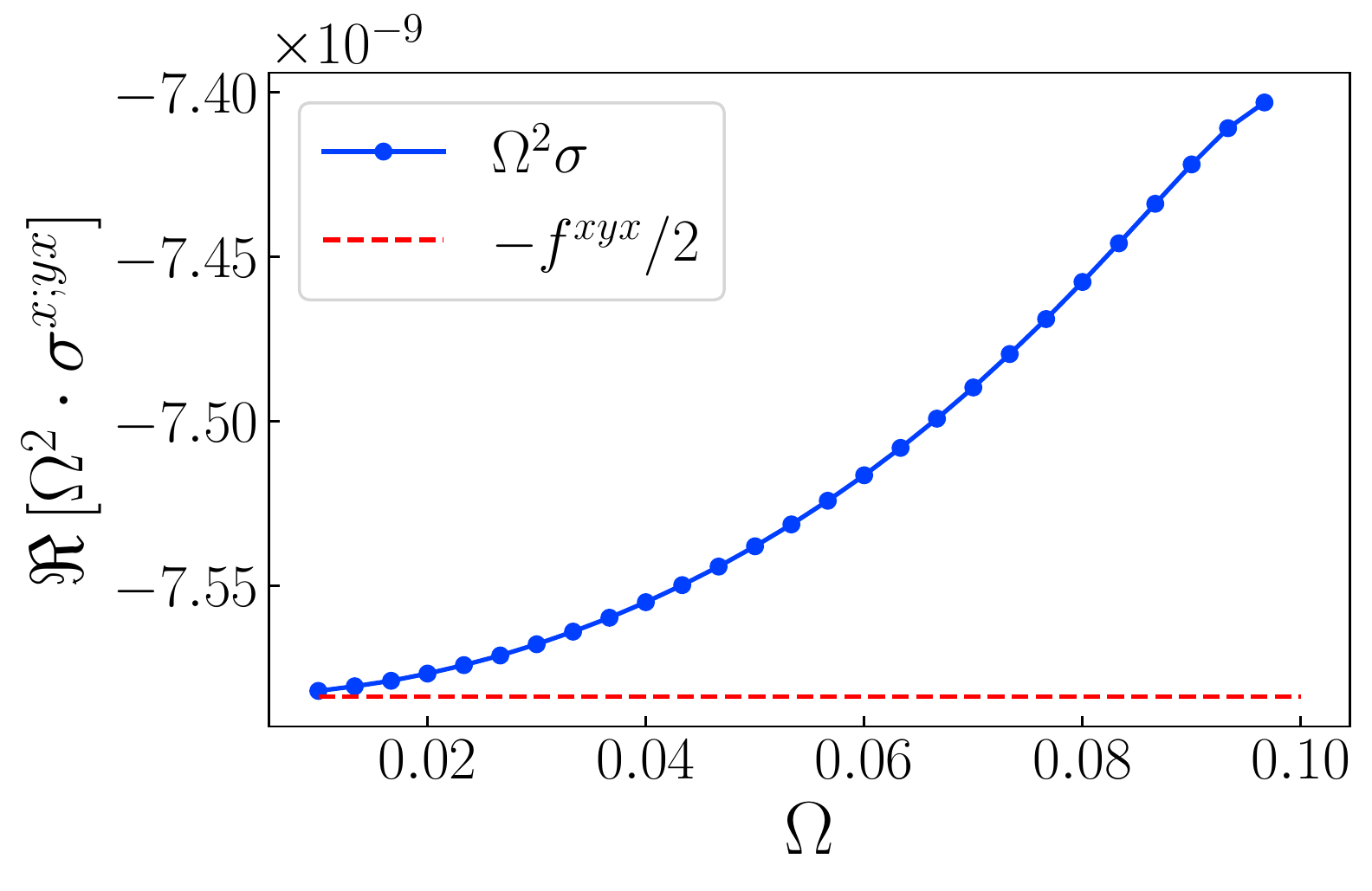}
        \caption{
        Comparison between the photocurrent conductivity and numerically calculated nonreciprocal superfluid density.
        The photocurrent conductivity $\Re{[\,\sigma^{x;yx}]}$ (blue solid line) and the nonreciprocal superfluid density $f^{xyx}$ (red dashed line) are calculated by Eqs.~\eqref{total_photocurrent_conductivity_gapped} and~\eqref{NSF_numerical_formula}, respectively.
        We take $T=10^{-4}$ and $N=3000$.
        The calculation of $\Re{[\,\sigma^{x;yx}]}$ is performed with $\gamma=10^{-3}$.
        }
        \label{Fig_NRSF_PGE_comparison}
        \end{figure}
\fi

\subsection{Anomalous contributions to NRO responses}
\label{Secsub_anomalous_PGE_SHG_model_study}

The divergent NRO responses are similarly observed in the other photocurrent conductivity components and in other NRO responses such as the second harmonic generation.
In this section, we show the frequency dependence of all the allowed components of the photocurrent and second harmonic generation coefficients.
We distinguish the two response functions by the second harmonic generation coefficient
$\sigma^{\alpha;\beta\gamma}_\text{SHG} =\sigma^{\alpha;\beta\gamma} (2\Omega;\Omega,\Omega)$ and the photocurrent conductivity $\sigma^{\alpha;\beta\gamma}_\text{PC} =\sigma^{\alpha;\beta\gamma} (0;\Omega,-\Omega)$.
The numerical calculations are performed with the total NRO conductivity formula [Eqs.~\eqref{RDM_2;0} to~\eqref{RDM_0;2_1photon_2}] where frequencies are taken by $\omega_1 = \pm \omega_2 =\Omega$.

First, we consider the \T{} symmetric parity-mixed superconductor.
The low-frequency divergence manifests in the imaginary part of the NRO responses [Figs.~\ref{Fig_T_all_NRO}(b) and \ref{Fig_T_all_NRO}(d)].
The divergence originates from the Berry curvature derivative $\hat{B}_d$.
In agreement with the analytical expression in Eq.~\eqref{general_static_conductivity_derivative}, $\Im{[\,\sigma_\text{SHG}^{x;xy}]} = \Im{[\,\sigma_\text{SHG}^{x;yx}]}$ and $\Im{[\,\sigma_\text{PC}^{x;xy}]}= -\Im{[\,\sigma_\text{PC}^{x;yx}]}$ converge to the same value in the low-frequency limit [see green-colored plots in Figs.~\ref{Fig_T_all_NRO}(b) and \ref{Fig_T_all_NRO}(d)].
$\Im{[\,\sigma_\text{SHG}^{y;yy}]}$ does not show divergence in the low-frequency regime, since the vanishing totally-antisymmetric tensor $\epsilon_{yy\delta}=0$ forbids the Berry curvature derivative effect. 
Note that the divergence does not appear in the real part of the NRO conductivity except for an artificial contribution arising from the phenomenological scattering rate as discussed in Sec.~\ref{Secsub_scattering_rate}.

        \expandafter\ifx\csname iffigure\endcsname\relax
        \begin{figure*}[htbp]
            \centering
            \includegraphics[width=0.70\linewidth,clip]{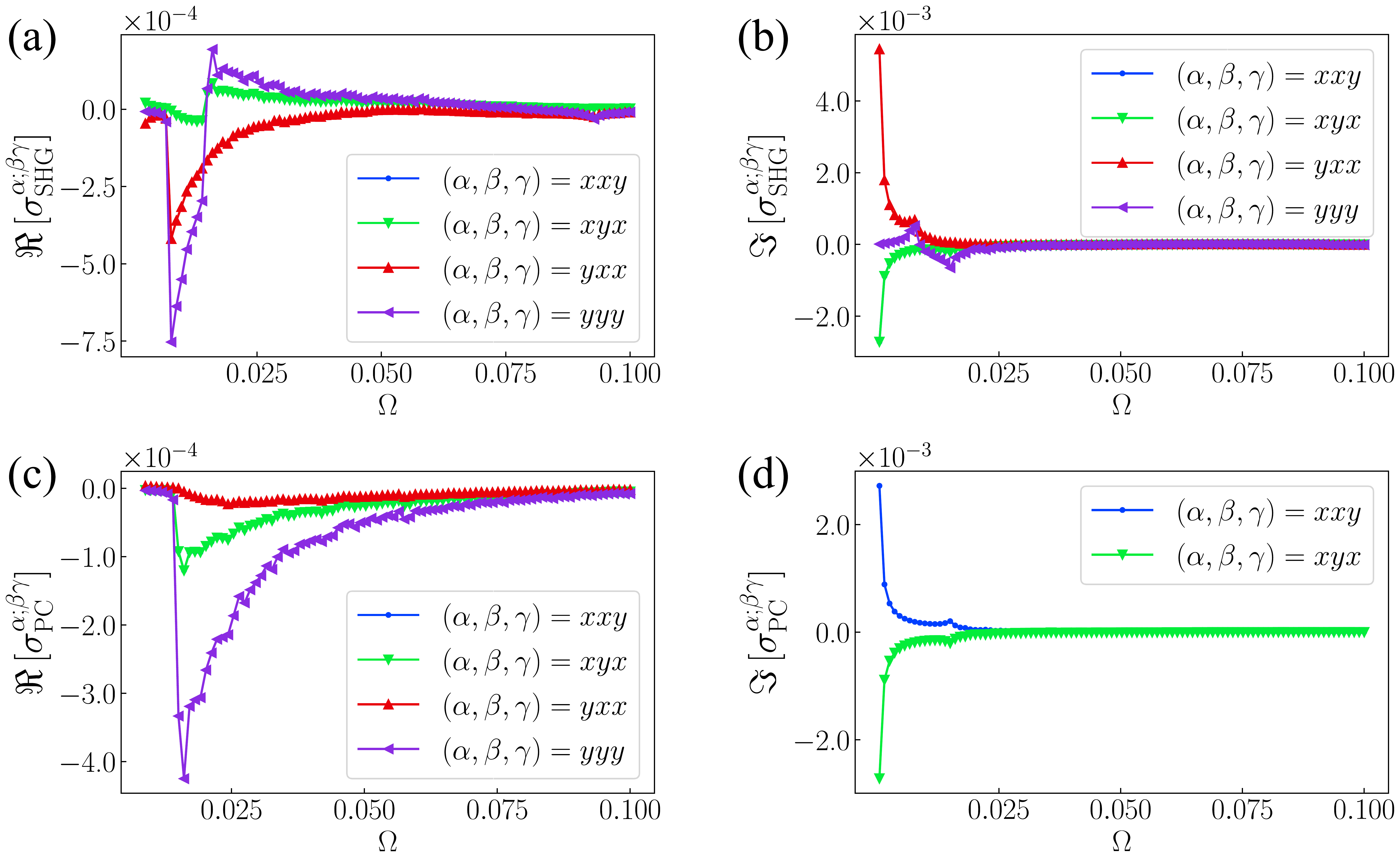}
            \caption{
            Frequency dependence of the second harmonic and photocurrent generation in a \T{} symmetric superconducting state.
            (a) $\Re{[\,\sigma_\text{SHG}^{\alpha;\beta\gamma}]}$ and (b) $\Im{[\,\sigma_\text{SHG}^{\alpha;\beta\gamma}]}$ for the second harmonic generation. 
            (c) $\Re{[\,\sigma_\text{PC}^{\alpha;\beta\gamma}]}$ and (d) $\Im{[\,\sigma_\text{PC}^{\alpha;\beta\gamma}]}$ for the photocurrent generation.
            We take $\gamma= 10^{-4}$, $T= 10^{-4}$, and $N=4000$.}
        \label{Fig_T_all_NRO}
        \end{figure*}
        \fi

Next, we look into the NRO response functions for the \PT{} symmetric parity-mixed superconductor.
The low-frequency divergence is observed in the real part of the NRO responses [Figs.~\ref{Fig_PT_all_NRO}(a) and \ref{Fig_PT_all_NRO}(c)] due to the nonreciprocal superfluid density $\hat{f}$.
Since $\hat{f}$ is the totally-symmetric tensor, $f^{xxy}=f^{yxx}$ leads to the same low-frequency divergence in $\Re{[\,\sigma^{x;xy}]}$ and $\Re{[\,\sigma^{y;xx}]}$, while $\Re{[\,\sigma_\text{SHG}^{x;xy}]} = \Re{[\,\sigma_\text{SHG}^{x;yx}]}$ and $\Re{[\,\sigma_\text{PC}^{x;xy}]} = \Re{[\,\sigma_\text{PC}^{x;yx}]}$ are satisfied by definition.

\expandafter\ifx\csname iffigure\endcsname\relax
\begin{figure*}[htbp]
    \centering
    \includegraphics[width=0.70\linewidth,clip]{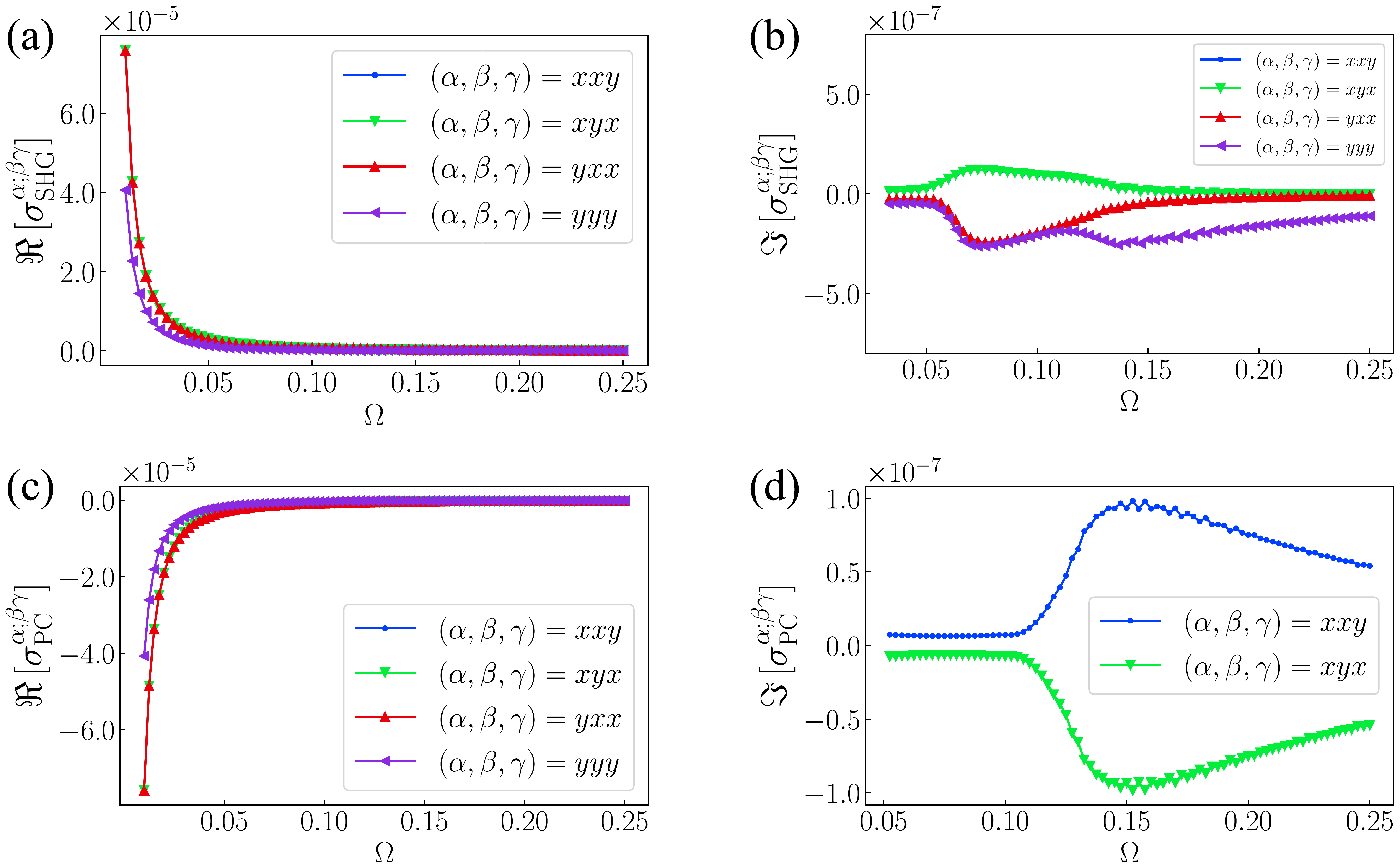}
    \caption{
    Frequency dependence of the second harmonic and photocurrent generation in a \PT{} symmetric superconducting state.
    (a) $\Re{[\,\sigma_\text{SHG}^{\alpha;\beta\gamma}]}$ and (b) $\Im{[\,\sigma_\text{SHG}^{\alpha;\beta\gamma}]}$ for the second harmonic generation. 
    (c) $\Re{[\,\sigma_\text{PC}^{\alpha;\beta\gamma}]}$ and (d) $\Im{[\,\sigma_\text{PC}^{\alpha;\beta\gamma}]}$ for the photocurrent generation.
    We take $\gamma= 3\times 10^{-3}$, $T= 10^{-4}$, and $N=4000$.}
\label{Fig_PT_all_NRO}
\end{figure*}
\fi

\subsection{Scattering rate dependence of anomalous NRO responses}
\label{Secsub_scattering_rate}

We address the scattering rate dependence of the anomalous NRO responses by taking the photocurrent conductivity as an example.

For the anomalous NRO responses in the low-frequency regime, numerical results show the diverging behavior.
$\Re{[\sigma^{x;yx}]}$ ($\Im{[\sigma^{x;yx}]}$) in the low-frequency regime is, however, influenced by the scattering rate for the \T{} symmetric (\PT{} symmetric) parity-breaking superconductor.
In contrast, $\Im{[\sigma^{x;yx}]}$ ($\Re{[\sigma^{x;yx}]}$) for the \T{} symmetric (\PT{} symmetric) parity-breaking superconductor is intrinsic and hence robust to scattering-rate effects.
These contrasting properties are revealed by calculating the photocurrent conductivity with scattering rates ranging over several orders of magnitude (Fig.~\ref{Fig_xyx_photocurrent_disorder_dependence}).
For instance, in Fig.~\ref{Fig_xyx_photocurrent_disorder_dependence}(a) for the \T{} symmetric case, $\Re{[\sigma^{x;yx}]}$ disappears with decreasing the scattering rate, whereas $\Im{[\sigma^{x;yx}]}$ shows the remarkable tolerance to the scattering effects.
According to Table~\ref{Table_photocurrent_classification}, the scattering-rate-sensitive part may be artificial or extrinsic.
Thus, we did not show such part in the previous subsections.

In the \T{} symmetric system, we do not have any diverging linearly-polarized-light-induced photocurrent response when we take into account scattering effects by implementing the Green function method [Eq.~\eqref{general_anomalous_divergent_NRO_conductivity}].
Thus, the low-frequency divergence of $\Re{[\sigma^{x;yx}]}$ is an artificial effect arising from the phenomenology of scattering effects.
On the other hand, a tiny circularly-polarized-light-induced photocurrent response in the \PT{} symmetric system [Fig.~\ref{Fig_xyx_photocurrent_disorder_dependence}(d)] may be an extrinsic effect due to the Drude derivative effect.
We can identify a genuine extrinsic contribution by treating the scattering effect in a rigorous way beyond the phenomenological scattering rate approximation.
Such rigorous calculation is left for future study.
The main finding of this numerical work is the intrinsic low-frequency divergence of the imaginary part of the NRO conductivity in the \T{} symmetric superconductor and the real part in the \PT{} symmetric superconductor, which are robust to scattering effects.
The scattering-tolerant property ensures that the Berry curvature derivative $\hat{B}_d$ and the nonreciprocal superfluid density $\hat{f}$ are promising indicators of the NRO phenomena in superconductors.

\expandafter\ifx\csname iffigure\endcsname\relax
        \begin{figure*}[htbp]
        \centering
        \includegraphics[width=0.70\linewidth,clip]{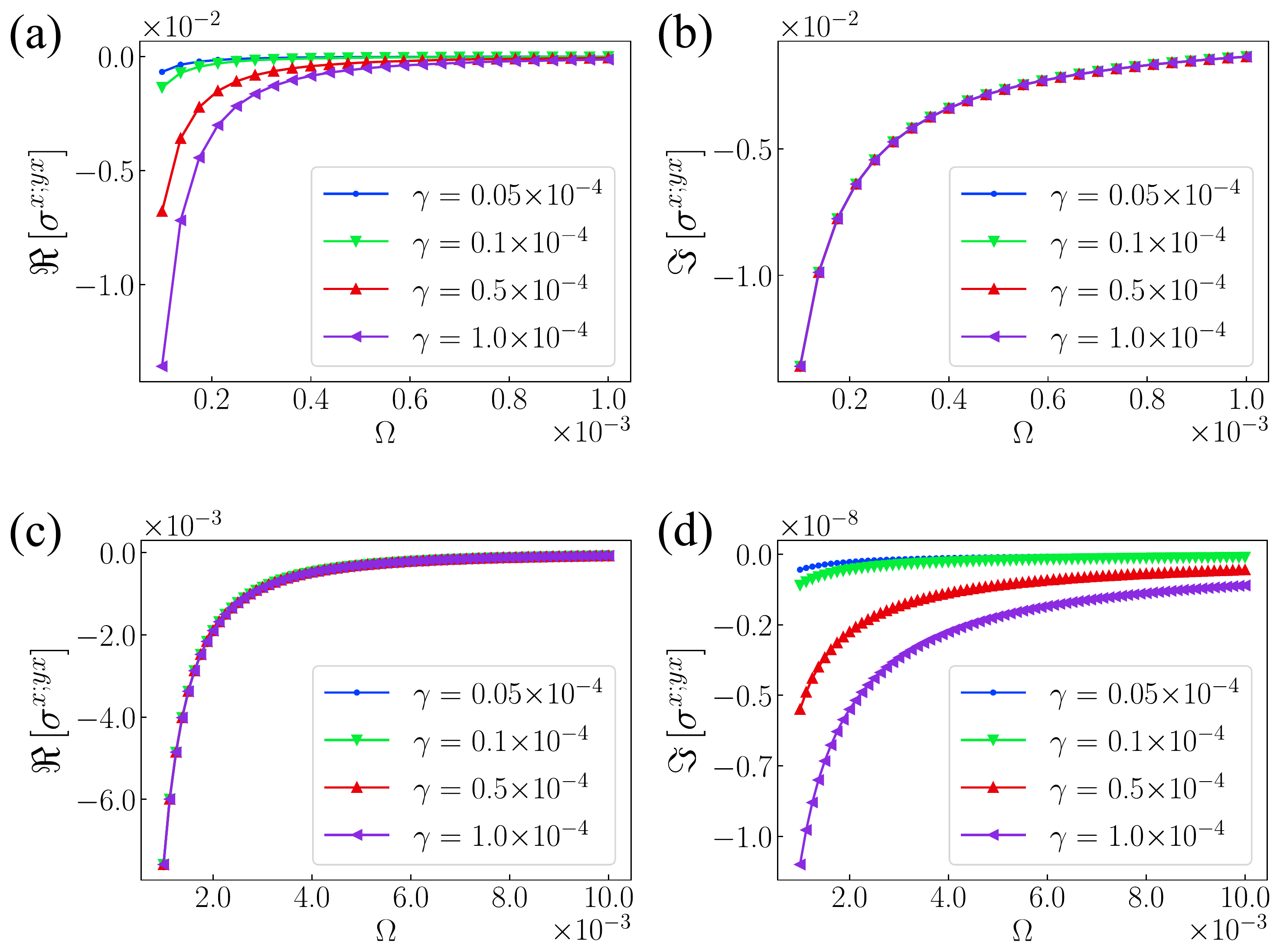}
            \caption{
            Scattering rate dependence of the total photocurrent conductivity $\sigma^{x;yx}$.
            Frequency dependence of the photocurrent conductivity is plotted for (a,\,b) the case of \T{} symmetric superconductors and for (c,\,d) the case of \PT{} symmetric superconductors.
            In each panel, a series of scattering rates are distinguished by color.
            In (b) and (c), all the plots are overlapped, indicating the tolerance to scattering effects.
            We take the temperature $T= 10^{-4}$ and $N=1500$.}
        \label{Fig_xyx_photocurrent_disorder_dependence}
        \end{figure*}
\fi

\section{Discussion and Summary}
\label{Sec_summary}

This paper elaborated the nonreciprocal optical responses characteristic to parity-breaking superconductors.
Based on the analytical and numerical calculations, we showed that the anomalous NRO responses play essential roles in superconductors.
The key ingredients, namely, nonreciprocal superfluid density and Berry curvature derivative, manifest in the superconducting state, although they have no counterpart in the normal state.
Since the anomalous NRO response is intrinsically divergent under the low-frequency light such as sub-terahertz light, it surpasses the NRO responses of other origins clarified in prior studies~\cite{Sturman1992Book}.
Our microscopic calculations confirmed that the anomalous NRO responses dominate the photocurrent and second harmonic generation responses.
For instance, for the linearly-polarized-light-induced photocurrent in the \PT{} symmetric parity-breaking superconductor, the normal photocurrent called injection current is negligible compared to the nonreciprocal superfluid density effect [Fig.~\ref{Fig_photocurrent_comparison_xyx}(c)].

A promising application of our work is a probe for superconducting symmetry.
The parity-mixing effect of the superconducting pairings has been drawing much attention in superconducting science from the discovery of noncentrosymmetric superconductors~\cite{NCSC_book}.
The NRO responses can appear in the parity-mixed superconductors and thus provide an estimate of the parity-mixing effect.
The estimate may be demonstrated by incorporating chemical or physical operation, which tunes the parity-mixing effect~\cite{Harada2012,Kitagawa2020}.
Furthermore, the anomalous NRO response is in close relation with the temporal symmetry preserved in superconductors.
Indeed, the photocurrent response induced by the circularly-polarized (linearly-polarized) light shows the characteristic divergence for the low-frequency light when the parity-breaking superconductor preserves \T{} (\PT{}) symmetry.
Recent studies implied the parity violation in exotic quantum materials such as cuprates~\cite{Zhao2017,De_la_Torre2020,Lim2020} and heavy-fermion systems~\cite{Braithwaite2019,Ishizuka2021}.
Thus, our classification of the NRO responses will be helpful for identifying the spatial and temporal symmetry of these materials.

It is also essential to explore novel functional devices utilizing the remarkable properties of the anomalous NRO response in superconductors.
Owing to the enhanced NRO response for the low-frequency light, the parity-breaking superconductor may realize a highly efficient optical apparatus for future terahertz light technology.
The anomalous NRO response arises from the condensed Cooper pairs, and hence the photo-electric conversion is expected to be carried out without undesirable Joule heating.
The photo-diode without energy loss possibly originates from the quantum nature of superconductivity.
Interestingly, the energy-saving NRO performance of superconductors can be turned on by injecting the supercurrent~\cite{Nakamura2020} [Case (II) in Sec.~\ref{Sec_symmetry_analysis}].
The tunability is a consequence of the superfluidity and cannot be found in systems without macroscopic quantum condensate.
The supercurrent-induced parity violation can be induced in various superconductors such as prototypical superconducting alloys and high-temperature copper-based superconductors.
As a result, superconductivity-based optoelectronics are available in a broad range of materials.

Our systematic study of the nonreciprocal optical response is expected to stimulate further investigations of superconductors in the scientific and engineering fields and make essential building blocks for superconductor-based optoelectronics.

\section*{Acknowledgements}
We thank Kyosuke Adachi, Shota Kanasugi, Hiroto Tanaka, and Yoshihiro Michishita for fruitful discussions.
H.W. is grateful to Naoto Nagaosa for enlightening comments.
This work was supported by JSPS KAKENHI (Grants No. JP18H05227, No. JP18H01178, and No. 20H05159), SPIRITS 2020 of Kyoto University.
H.W. is a JSPS research fellow and supported by JSPS KAKENHI (Grant No.~18J23115 and No.~21J00453).
A.D. is supported by JSPS KAKENHI (Grant No.~21K13880).

\clearpage
\appendix
\onecolumngrid

\section{Density matrix theory based on Bogoliubov-de Gennes Hamiltonian}
\label{App_Sec_density_matrix_formulation}

This section is devoted to the density matrix formulation on the basis of the BdG Hamiltonian.
We note that some expressions are overlapped with those in the main text.

The Hamiltonian is given by the BdG form,
		\begin{equation}
        \mathcal{H}_\text{BdG} = \frac{1}{2} \bm{\Psi}^\dagger H_\text{BdG}\bm{\Psi} +\text{const.},\label{eq:BdGtemp}
        \end{equation}
where Nambu spinor $\bm{\Psi}^\dagger = (\bm{c}^\dagger,\bm{c}^T)$ with the creation ($\bm{c}^\dagger$) and annihilation ($\bm{c}$) operators of electrons.
Here, the components of $\bm{c}$ are the normal-state degrees of freedom, namely, spatial position, sublattice, orbital, spin, and so on.
The BdG Hamiltonian $H_{\text{BdG}}$ includes the Hamiltonian in the normal state $H_\text{N}$ and the pair potential of superconductivity $\Delta$
\begin{equation}
    H_{\text{BdG}}(\bm{A})=\begin{pmatrix}H_\text{N}(\bm{A})&\Delta\\\Delta^\dagger&-H_\text{N}(\bm{A})^T\end{pmatrix}.
    \label{eq:BdG_A}
\end{equation}
The electromagnetic perturbation is taken into account by the vector-potential $\bm{A}$. In equilibrium,  we take $\bm{A}=0$.

Owing to the particle-hole symmetry of $H_\text{BdG}$, the spinor satisfies the relation
            \begin{equation}
            [\mathcal{H}_\text{BdG}, \Psi_{i}] = - \left(  H_\text{BdG}\right)_{ij} \Psi_{j}.
            \end{equation}
Here, the index $i,j$ run over both the normal and the Nambu degrees of the freedom.
    
It is convenient to retake the basis of the Nambu spinor $\Psi_i$ to the energy eigenstates of $H_{\text{BdG}}$,
$[ H_\text{BdG},\Phi_{a}] = -E_{a}\Phi_{a} $.
This commutation relation indicates that the density-matrix formalism straightforwardly applies to the BdG Hamiltonian as follows.
We introduce the density matrix $\hat{\rho}(t)$ as in the normal state.
Without perturbation, it is written as
\begin{equation}
    \hat{\rho}^{(0)}=e^{-\beta\mathcal{H}_{\text{BdG}}}/\Tr[e^{-\beta\mathcal{H}_{\text{BdG}}}],
\end{equation}
and the time evolution of $\hat{\rho}(t)$ is described by the von-Neumann equation
\begin{equation}
    i\partial_t\hat{\rho}(t)=[\mathcal{H}_{\text{BdG}}+\Delta\mathcal{H}(t),\hat{\rho}(t)].\label{von_Neumann_equation_formal}
\end{equation}
Here, $\Delta \mathcal{H} (t) = \bm{\Psi}^\dagger \Delta H (t) \bm{\Psi}/2$ is the perturbative Hamiltonian.
It is convenient to introduce the reduced density matrix described by the Bogoliubov quasiparticles
    \begin{equation}
    \rho_{ab} (t) = \Tr [\Phi_b^\dagger \Phi_a \hat{\rho} (t)],
    \end{equation}
and accordingly obtain the von-Neumann equation
    \begin{align}
    i \partial_t \rho_{ab} =E_{ab}\rho_{ab}(t)+  \left[ \Delta H (t), {\rho }(t)\right]_{ab}, \label{von_Neumann_equation}
    \end{align}
where
$E_{ab}= E_a - E_b$ and $[A,B]_{ab} = A_{ac}B_{cb}-A_{cb}B_{ac}$. In equilibrium, the density matrix $\rho_{ab}^{(0)} (t) = f_a \delta_{ab} $ is reduced to the Fermi-Dirac distribution function, $f_a\equiv(e^{\beta E_a}+1)^{-1}$. 
Throughout this work, we consider the photoelectric field expressed in the velocity gauge. The perturbative Hamiltonian is given by $\Delta H(t)=H_{\mathrm{BdG}}(\bm{A}(t))-H_{\mathrm{BdG}}(\bm{0})$, that is,
		\begin{equation}
        \Delta H (t) = \sum_{n=1} \frac{1}{n!} \left(  -1\right)^n A^{\alpha_1} (t) \cdots A^{\alpha_n} (t) \vj^{\alpha_1 \cdots \alpha_n},\label{perturbative_Hamiltonian}
        \end{equation}
where we introduce the generalized velocity operator 
    \begin{equation}
    \vj^{\alpha_1 \cdots \alpha_n} =(-1)^n \frac{\partial^n H_\text{BdG}(\bA)}{\partial A^{\alpha_1}\cdots \partial A^{\alpha_n}}\Bigr|_{\bA=0}.\label{generalized_velocity_operator}
    \end{equation}
For the BdG Hamiltonian $H_\text{BdG}(\bA)$, the vector potential is coupled to the diagonal part in the form of the minimal coupling [Eq.~\eqref{eq:BdG_A}].
Note that we do not take into account field-induced corrections to the pair potential for simplicity.
Integrating Eq.~\eqref{von_Neumann_equation} iteratively, we obtain the perturbed density matrix $ \hat{\rho} (t)$.
Then, we immediately obtain the expectation value of the electric current as
		\begin{equation}
        \Braket{\mathcal{J}^\alpha}(t) = \Tr [ \hat{\rho} (t) \mathcal{J}^\alpha (t) ], 
        \end{equation}
with the velocity operator 
		\begin{equation}
            \mathcal{J}^\alpha (t) = \frac{1}{2}\bm{\Psi}^\dagger \vj^\alpha_\bA (t) \bm{\Psi}. \label{velocity_operator_BdGform}
        \end{equation}
The matrix denote the modified velocity operator
		\begin{equation}
            \vj_\bA^\alpha (t) = \sum_{m=0} \frac{1}{m!} \left(  -1\right)^{m} A^{\beta_1} (t) \cdots A^{\beta_{m}} (t) \vj^{\alpha\, \beta_1 \cdots \beta_m}.\label{perturbative_velocity_operator}
        \end{equation}

Here, we give the perturbed density matrix by expanding the density matrix in Eq.~\eqref{von_Neumann_equation} as 
    \begin{equation}
    \rho_{ab} (t) = \sum_{n=0} \rho_{ab}^{(n)}(t).
    \end{equation}
$\rho_{ab}^{(n)}$ denotes the $O(|\bm{A}|^n)$ correction to the density matrix. Accordingly, the perturbed density matrix up to the second-order are obtained as
    \begin{align}
    &\left( i\partial_t -E_{ab} \right) \rho_{ab}^{(0)} (t)=0,\\
    &\left( i\partial_t -E_{ab} \right) \rho_{ab}^{(1)} (t)=- A^{\alpha} (t) [ \vj^{\alpha}, \rho^{(0)} (t)]_{ab},\\
    &\left( i\partial_t -E_{ab} \right) \rho_{ab}^{(2)} (t)=-  A^{\alpha} (t) [ \vj^{\alpha}, \rho^{(1)} (t)]_{ab}+ \frac{1}{2} A^{\alpha} (t) A^{\beta}(t) [ \vj^{\alpha\beta}, \rho^{(0)} (t)]_{ab}.
    \end{align}
Taking the convention for Fourier transformation as
    \begin{equation}
    X (t) = \int_{-\infty}^{\infty} \frac{d\omega}{2\pi} X(\omega) e^{-i\omega t-\eta t},
    \end{equation}
we obtain the perturbed density matrix in the frequency domain. The adiabaticity parameter $\eta$ is positive infinitesimal. The density matrix are $\rho_{ab}^{(0)} (\omega ) = f_a \delta_{ab}  ~2\pi  \delta (\omega)$ and 
    \begin{align}
    \rho_{ab}^{(1)} (\omega)
        &=- \frac{1}{\omega + i \eta -E_{ab} }  A^{\alpha} (\omega) \vj^{\alpha}_{ab}f_{ba},  \label{RDM_1st_fourier}\\
    \rho_{ab}^{(2)} (\omega) 
        &= \frac{1}{ \omega + 2i\eta -E_{ab}} \int \frac{d\omega_1 d\omega_2 }{(2\pi)^2} (2\pi) \delta (\omega -\omega_1 - \omega_2)\notag  \\
        &\times \left( -   A^{\alpha}(\omega_1) [ \vj^{\alpha}, \rho^{(1)} (\omega_2)]_{ab}  + \frac{1}{2}  A^{\alpha} (\omega_1) A^{\beta} (\omega_2) \vj^{\alpha\beta}_{ab} f_{ba}\right). \label{RDM_2nd_fourier}
    \end{align}

The expectation value of the electric current is calculated by the current operator [Eq.~\eqref{generalized_velocity_operator}] and the perturbed density matrix, 
    \begin{equation}
    2 \Braket{\mathcal{J}^\alpha (t)} = 2\Tr{[\mathcal{J}^\alpha (t) \hat{\rho} (t)]} =  \sum_{a,b}\vj_{ab}^\alpha (t)\rho_{ba} (t).
    \end{equation}
    The perturbed electric current operator reads
    \begin{align}
        \mathcal{J}^\alpha(t)
            &=\mathcal{J}_{(0)}^\alpha(t)+\mathcal{J}_{(1)}^\alpha(t)+\mathcal{J}_{(2)}^\alpha(t)+\cdots,\\
        \mathcal{J}_{(0)}^\alpha(t)
            &=\int\frac{d\omega}{2\pi}e^{-i\omega t}\,2\pi\delta(\omega)\mathcal{J}^\alpha,\\
        \mathcal{J}_{(1)}^\alpha(t)
            &=-\int\frac{d\omega}{2\pi}e^{-i\omega t+\eta t}\mathcal{J}^{\alpha\beta}A^\beta(\omega),\\
        \mathcal{J}_{(2)}^\alpha(t)
            &=\int\frac{d\omega}{2\pi}e^{-i\omega t+2\eta t}\int\frac{d\omega_1d\omega_2}{2\pi} \delta(\omega_1+\omega_2-\omega)\frac{1}{2}\mathcal{J}^{\alpha\beta\gamma}A^\beta(\omega_1)A^\gamma(\omega_2).
    \end{align}
Before proceeding to the second-order nonlinear optical conductivity, let us consider the linear optical conductivity. The linear response is given by 
    \begin{align}
    \Braket{2 \mathcal{J}^\alpha (\omega)}_{(1)}
        &= \int dt\ e^{-i\omega t-\eta  t} \Tr{[ 2\mathcal{J}^\alpha_{(1)}(t) \hat{\rho}^{(0)} (t) + 2\mathcal{J}^\alpha_{(0)}(t) \hat{\rho}^{(1)} (t) ]},\\
        &= \sum_{a,b}\left(   \frac{- \vj^\alpha_{ab} \vj^\beta_{ba}f_{ab}}{ \omega + i\eta -E_{ba}} A^\beta (\omega)  -  A^\beta (\omega) \vj^{\alpha\beta}_{ab}f_{a} \delta_{ab}  \right).
      \end{align}
In the last line, the former and the latter originate from the correction to the density matrix and current operator, which are so-called paramagnetic and diamagnetic contributions, respectively. With the velocity gauge $\bm{E} = -\partial_t \bm{A} (t)$, the linear optical conductivity is obtained as
    \begin{equation}
    2 \sigma^{\alpha\beta} (\omega) = \frac{1}{i\omega-\eta}\sum_{a,b}\left(   \frac{- \vj^\alpha_{ab} \vj^\beta_{ba}f_{ab}}{ \omega + i\eta -E_{ba}}   -  \vj^{\alpha\beta}_{ab}f_{a} \delta_{ab}  \right).\label{linear_conductivity}
    \end{equation}

Similarly, the second-order electic current is divided into three components,
		\begin{align}
        \Braket{\mathcal{J}^\alpha (\omega)}_{(2)}
            &= \int dt\ e^{-i\omega t-\eta  t}\Tr{[ \mathcal{J}^\alpha_{(2)}(t) \hat{\rho}^{(0)} (t) + \mathcal{J}^\alpha_{(1)}(t) \hat{\rho}^{(1)} (t) + \mathcal{J}^\alpha_{(0)}(t) \hat{\rho}^{(2)} (t)]}.\label{2nd_current_RDM}
        \end{align}
The second-order optical conductivity is explicitly given by
        \begin{align}
            & 2\sigma^{\alpha;\beta\gamma} (\omega;\omega_1,\omega_2) \nonumber \\
                & =\frac{1}{i\omega_1-\eta}\frac{1}{i\omega_2-\eta}  \left[ \sum_a \frac{1}{2}  \vj^{\alpha\beta\gamma}_{aa} f_a \label{App_RDM_2;0} \right.\\ 
                &\left.+\sum_{a,b} \frac{1}{2} \left(  \frac{\vj^{\alpha\beta}_{ab} \vj^\gamma_{ba} f_{ab} }{ \omega_2 +i\eta -E_{ba}} +  \frac{\vj^{\alpha\gamma}_{ab} \vj^\beta_{ba} f_{ab} }{ \omega_1 +i\eta -E_{ba}}\label{App_RDM_1;1}\right)\right. \\
                &\left.+\sum_{a,b}\frac{1}{2}  \frac{\vj^{\alpha}_{ab} \vj^{\beta\gamma}_{ba} f_{ab} }{ \omega + 2i\eta -E_{ba}}\label{App_RDM_0;2_2photon}\right.\\
                &\left.+\sum_{a,b,c} \frac{1}{2}\frac{\vj^{\alpha}_{ab}}{ \omega + 2i\eta -E_{ba}} \left(   \frac{\vj^\beta_{bc}\vj^{\gamma}_{ca}f_{ac} }{\omega_2 + i \eta -E_{ca} }    -    \frac{\vj^\beta_{ca}\vj^{\gamma}_{bc}f_{cb} }{\omega_2 + i \eta -E_{bc} }  \right)\label{App_RDM_0;2_1photon_1}\right.\\
                &\left.+\sum_{a,b,c}\frac{1}{2}\frac{\vj^{\alpha}_{ab}}{ \omega + 2i\eta -E_{ba}} \left(  \frac{\vj^\gamma_{bc}\vj^{\beta}_{ca}f_{ac} }{\omega_1 + i \eta -E_{ca} }    -    \frac{\vj^\gamma_{ca}\vj^{\beta}_{bc}f_{cb} }{\omega_1 + i \eta -E_{bc} } \right)\right], \label{App_RDM_0;2_1photon_2}
        \end{align} 
in which the first line, second line, and third to fifth lines correspond to the first, second, and third components in Eq.~\eqref{2nd_current_RDM}, respectively.
Note that only Eqs.~\eqref{App_RDM_0;2_1photon_1} and~~\eqref{App_RDM_0;2_1photon_2} appear in electron systems in the continuum space~\cite{Von_Baltz1981}.
We confirmed that the obtained formulas are consistent with those in the normal state~\cite{Passos2018,Michishita2020,Gao2020Photo} by replacing the vector potential with the crystal momentum $\bk$ and by ignoring the pair potential.

\section{Phenomenological scattering rate}
\label{App_Sec_Phenomenology}

Here, we introduce the phenomenological treatment of the scattering rate. The scattering rate $\gamma$ is introduced by modifying the von-Neumann equation in Eq.~\eqref{von_Neumann_equation_formal} to 
\begin{equation}
    i\partial_t \hat{\rho}^{(n)}  =\sum_{m=0}^n \left[ \mathcal{H}^{(n-m)} (t), \hat{\rho}^{(m)} (t)  \right] - i n \gamma \hat{\rho}^{(n)}, \label{von_Neumann_equation_with_scattering}
\end{equation}
where $\hat{\rho}^{(n)}$ and $\mathcal{H}^{(n)}$ are $O(F^n)$ where $F$ represents the strength of the external field. The scattering rate is therefore multiplied by the perturbation order $n$ of the density matrix. This description is reasonable to obtain physically meaningful results. Implementing the phenomenology in Eq.~\eqref{von_Neumann_equation_with_scattering}, Ref.~\cite{Passos2018} obtained the (third-order) nonlinear optical conductivity spectrum consistent with the gauge symmetry. 

Following Eq.~\eqref{von_Neumann_equation_with_scattering}, we obtain the formula for the nonreciprocal optical conductivity by replacing the adiabaticity parameter $\eta$ with $\gamma$ in Eqs.~\eqref{RDM_2;0}-\eqref{RDM_0;2_1photon_2}. Note that the prefactor in Eq.~\eqref{RDM_2;0}, $(i\omega_1 -\eta)^{-1}(i\omega_2 -\eta)^{-1}$, does not include the scattering rate $\gamma$ since it appears by converting the vector potential into the electric field.

\section{Linear optical conductivity}
\label{App_Sec_linear_optical_conductivity}

We demonstrate that $\bm{\lambda}$-parametrization introduced in Sec.~\ref{Secsub_setup} leads to a physically-transparent expression by investigating the linear conductivity in Eq.~\eqref{linear_conductivity}. Similarly to Eq.~\eqref{hellmann_feynman_velocity_vector}, the diamagnetic current operator obeys the relation
		\begin{equation}
                  \vj^{\alpha\beta}_{aa}= \braket{a| \partial_{\lambda_\alpha}\partial_{\lambda_\beta}H|a}= -\partial_{\lambda_\beta} \vj^\alpha_{aa} + i \left[ \xi^{\lambda_\beta}, \vj^\alpha \right]_{aa},  \label{hellmann_feynman_diamagnetic_current_vector}
        \end{equation}
where $\bm{\lambda}$ dependence is suppressed. The linear conductivity in Eq.~\eqref{linear_conductivity} is therefore simplified as
    \begin{align}
    2\sigma^{\alpha\beta} (\omega)
        &= \frac{1}{i\omega-\eta}\left[ \sum_{a} - \partial_{\lambda_\alpha} \partial_{\lambda_\beta} E_af_{a}  - \sum_{a\neq b} \left( \frac{- \vj^\alpha_{ab} \vj^\beta_{ba}f_{ab}}{ \omega + i\eta -E_{ba}} - i \xi^{\lambda_\beta}_{ab}\vj^\alpha_{ba}f_{ab} \right) \right] ,\\
        &= \frac{1}{i\omega-\eta}\left[ \sum_{a} - \partial_{\lambda_\alpha} \partial_{\lambda_\beta} E_af_{a}  + \sum_{a\neq b} \vj^\alpha_{ab} \vj^\beta_{ba}f_{ab} \left(  \frac{1}{ \omega + i\eta -E_{ba}} - \frac{1}{E_{ab}}  \right) \right] , \\
        &= \frac{1}{i\omega-\eta} \left[\sum_{a} - \partial_{\lambda_\alpha} \partial_{\lambda_\beta} E_af_{a}  - (  \omega + i\eta ) \sum_{a\neq b} \frac{\vj^\alpha_{ab} \vj^\beta_{ba}f_{ab}}{( \omega + i\eta -E_{ba}) E_{ab}}\right].
        \label{eq:linear_conductivity}
      \end{align}
From the obtained expression, we see the clear physical meaning of each contribution to the linear optical conductivity. For instance, the second term in the right hand side represents an interlevel transition process and is rewritten by
		\begin{align}
            i \sum_{a\neq b} \frac{\vj^\alpha_{ab} \vj^\beta_{ba}f_{ab}}{( \omega + i\eta -E_{ba}) E_{ab}}= -i \sum_{a\neq b} \frac{E_{ab}}{ \omega + i\eta -E_{ba}} \left(  g^{\lambda_\alpha \lambda_\beta}_{ab} -\frac{i}{2}\Omega^{\lambda_\alpha \lambda_\beta}_{ab} \right)  f_{ab},
        \end{align}
where we introduced the band-resolved quantum metric and Berry curvature~\cite{Gao2020,Watanabe2021} 
		\begin{equation}
            g^{\lambda_\alpha \lambda_\beta}_{ab} = \Re{[\xi^{\lambda_\alpha}_{ab}\xi^{\lambda_\beta}_{ba}]},~\Omega^{\lambda_\alpha \lambda_\beta}_{ab} = -2\Im{[\xi^{\lambda_\alpha}_{ab}\xi^{\lambda_\beta}_{ba}]}. \label{quantum_metric_and_Berry_curvature_in_vector_potential}
        \end{equation} 
In this way, this term has a geometric meaning. In particular, taking the static limit $\omega \rightarrow 0$, the interlevel transition process is given by the Berry curvature
		\begin{equation}
        \sum_{b(\neq a)}  \Omega^{\lambda_\alpha \lambda_\beta}_{ab}   f_{a} = \epsilon_{\alpha\beta\gamma}  \Omega^{\lambda_\gamma}_{a}   f_{a}.
        \end{equation}
We introduce the Berry curvature parametrized by the variational parameter $\bm{\lambda}$
		\begin{equation}
            \Omega^{\lambda_\gamma}_a = \frac{1}{2}\sum_{b(\neq a)} \epsilon_{\alpha\beta\gamma} \Omega^{\lambda_\alpha \lambda_\beta}_{ab}. \label{charged_Berry_curvature}
        \end{equation}
Replacing $\bla$ with $\bk$, we can see that the obtained expression reproduces the formula for the anomalous Hall conductivity in the normal state.
Note that the Berry curvature effect does not lead to the quantized Hall conductivity in the superconducting state, since the variational parameter has no periodicity in terms of $2\pi/L$ ($L$ is the system size). In the normal state, the recovered periodicity of $\bk (\bla)$ leads to the quantized Hall conductivity in the gapped phase at the zero temperature~\cite{Niu1984}.
Note also that the Berry curvature defined in Eq.~\eqref{charged_Berry_curvature} is different from the Berry curvature contributing to the thermal Hall effect in superconductors~\cite{Nomura2012,Sumiyoshi2013}.
The latter one (``thermal'' Berry curvature) is defined by the connection Eq.~\eqref{eq:connection_temp}, and treats electrons and holes on equal footing, giving rise to the thermal transport.
On the other hand, Eq.~\eqref{charged_Berry_curvature} (``charge'' Berry curvature) is defined by Eq.~\eqref{eq:connection_temp2}, and includes electrons and holes with different signs, contributing to the charge transport as is derived here.

The remaining term in the linear conductivity [Eq.~\eqref{eq:linear_conductivity}] is the intralevel transition process denoted by
		\begin{equation}
            \frac{1}{i\omega-\eta} \sum_{a} - \partial_{\lambda_\alpha} \partial_{\lambda_\beta} E_af_{a}.\label{linear_conductivity_horizontal}
        \end{equation}
Recalling the Drude conductivity in the normal state, we can deduce the Drude part in Eq.~\eqref{linear_conductivity_horizontal}. Conducting the partial integration, we obtain
		\begin{equation}
            \frac{1}{i\omega-\eta} \sum_{a} \left[ - \partial_{\lambda_\beta} \left( \partial_{\lambda_\alpha}  E_a f_a \right) +  \partial_{\lambda_\alpha}  E_a  \partial_{\lambda_\beta} f_{a} \right].
        \end{equation}
The second term gives the Drude conductivity, whereas the first term should vanish in the normal state due to the periodicity of the Brillouin zone. On the other hand, the disappearance of the first term does not hold for the superconducting state. With the free energy calculated with the Hamiltonian $H_\bla$, 
        \begin{align}
            \sum_a\partial_{\lambda_\alpha}E_a f_a&=\Tr[(\partial_{\lambda_\alpha}H)f(H_\bla)]\\
            &=-\frac{1}{\beta}\partial_{\lambda_\alpha}\Tr\ln(1+e^{-\beta H_\bla})\\
            &=2\partial_{\lambda_\alpha} F_\bla.
        \end{align}
The prefactor 2 comes from the doubling of particles and holes.
Accordingly, we obtain
		\begin{align}
        \frac{-1}{i\omega-\eta} \sum_{a} \partial_{\lambda_\beta} \left( \partial_{\lambda_\alpha}  E_a f_a \right)
        &=\frac{-2}{i\omega-\eta} \partial_{\lambda_\alpha} \partial_{\lambda_\beta} F_\bla,\\
        &=\frac{-2}{i\omega-\eta} \rho^{\alpha\beta}_\text{s},\\
        &=-2 \rho^{\alpha\beta}_\text{s} \left(  \text{P}\frac{1}{i\omega} - \pi \delta (\omega)  \right),\label{App_superfluid_density}
        \end{align}
where the superfluid density is defined by $\rho^{\alpha\beta}_\text{s}=\partial_{\lambda_\alpha}\partial_{\lambda_\beta}F_\bla$.
In the final line, P denotes the principal integral for the frequency $\Omega$.
Finally, taking the limit $\bm{\lambda}\rightarrow \bm{0}$, we obtain the real part,
		\begin{equation}
        \pi \rho^{\alpha\beta}_\text{s} \delta (\omega),\label{linear_Meissner}
        \end{equation}
which ensures the zero electrical resistivity.
As a result, we reproduced the linear optical conductivity in the superconducting state~\cite{Tinkham1959SCskin,Tinkham2004introduction} based on the formulation using the variational parameter $\bm{\lambda}$.

\section{Derivation of formulas for photocurrent response}
\label{App_Sec_derivation_photocurrent_response}

The ${\bm \lambda}$-parametrization works in the formulation of NRO conductivity in superconductors as in the case of the linear optical conductivity (Appendix~\ref{App_Sec_linear_optical_conductivity}).
For example, we study the photocurrent response with the condition $\omega_1 = - \omega_2 = \Omega$ in Eqs.~\eqref{RDM_2;0} to~\eqref{RDM_0;2_1photon_2} and suppress the frequency dependence in the following.

The entire expression is given by
		\begin{equation}
        \sigma^{\alpha;\beta\gamma} = \sigma^{\alpha;\beta\gamma}_a + \sigma^{\alpha;\beta\gamma}_b + \sigma^{\alpha;\beta\gamma}_c + \sigma^{\alpha;\beta\gamma}_d,\label{total_photocurrent_conductivity}
        \end{equation}
where each contribution is given by
    \begin{align}
    2\sigma^{\alpha;\beta\gamma}_a
        &=\frac{1}{(i\Omega+i\delta - \eta)(-i\Omega+i\delta - \eta)}  \sum_a \frac{1}{2}  \vj^{\alpha\beta\gamma}_{aa} f_a \label{RDM_2;0_photocurrent}, \\ 
    2\sigma^{\alpha;\beta\gamma}_b
        &=\frac{1}{(i\Omega+i\delta - \eta)(-i\Omega+i\delta - \eta)} \sum_{a,b} \left( \frac{1}{2} \frac{\vj^{\alpha\beta}_{ab} \vj^\gamma_{ba} f_{ab} }{- \Omega + \delta +i\eta -E_{ba}} + \frac{1}{2} \frac{\vj^{\alpha\gamma}_{ab} \vj^\beta_{ba} f_{ab} }{ \Omega + \delta +i\eta -E_{ba}} \right) \label{RDM_1;1_photocurrent}, \\
    2\sigma^{\alpha;\beta\gamma}_c
        &= \frac{1}{(i\Omega+i\delta - \eta)(-i\Omega+i\delta - \eta)} \sum_{a,b}\frac{1}{2}  \frac{\vj^{\alpha}_{ab} \vj^{\beta\gamma}_{ba} f_{ab} }{2  \delta  + 2i\eta -E_{ba}},\label{RDM_0;2_2photon_photocurrent}\\
    2\sigma^{\alpha;\beta\gamma}_d
        &= \frac{1}{(i\Omega+i\delta - \eta)(-i\Omega+i\delta - \eta)} \notag \\
        &\times \sum_{a,b,c} \Biggl[ \frac{1}{2}\frac{\vj^{\alpha}_{ab}}{2 \delta  + 2i\eta -E_{ba}} \left(   \frac{\vj^\beta_{bc}\vj^{\gamma}_{ca}f_{ac} }{-\Omega+ \delta  + i \eta -E_{ca} }    -    \frac{\vj^\beta_{ca}\vj^{\gamma}_{bc}f_{cb} }{-\Omega+ \delta  + i \eta -E_{bc} }  \right)\notag \\
        &~~~~+\sum_{a,b,c}\frac{1}{2}\frac{\vj^{\alpha}_{ab}}{ 2 \delta + 2i\eta -E_{ba}} \left(  \frac{\vj^\gamma_{bc}\vj^{\beta}_{ca}f_{ac} }{\Omega+ \delta  + i \eta -E_{ca} }    -    \frac{\vj^\gamma_{ca}\vj^{\beta}_{bc}f_{cb} }{\Omega+ \delta  + i \eta -E_{bc} } \right)  \Biggr]. \label{RDM_0;2_1photon_photocurrent}
  \end{align}
For a technical reason, we introduce a finite sum-frequency $2\delta = \omega_1 + \omega_2 $~\cite{Sipe2000,De_Juan2020} and will take the limit $\delta \rightarrow 0$.

First, we take the component $\sigma_d$. When $E_a = E_b$, the expression diverges in the limit $\delta \rightarrow 0$ due to $E_{ab}=0$. Thus, we perform the expansion of what is enclosed in the parenthesis for $\delta$. For instance, taking the first term in Eq.~\eqref{RDM_0;2_1photon_photocurrent},  
            \begin{align}
            &\frac{1}{2}\frac{\vj^{\alpha}_{aa}\vj^\beta_{ac}\vj^{\gamma}_{ca}f_{ac} }{2 \delta  + 2i\eta } \left(   \frac{1}{-\Omega + \delta  + i \eta -E_{ca} } \right)\notag \\ 
                &= \frac{1}{2}\frac{\vj^{\alpha}_{aa}\vj^\beta_{ac}\vj^{\gamma}_{ca}f_{ac} }{2 \delta  + 2i\eta } \left(   \frac{1}{-\Omega  + i \eta -E_{ca}} - \frac{ \delta}{(-\Omega  + i \eta -E_{ca})^2}   +O(\delta^2) \right),\\  
                &\rightarrow \frac{1}{2} \vj^{\alpha}_{aa}\vj^\beta_{ac}\vj^{\gamma}_{ca}f_{ac} \left(   \frac{1}{ 2i\eta  (-\Omega  + i \eta -E_{ca}) } - \frac{1/2}{(-\Omega  + i \eta -E_{ca})^2}  \right). 
        \end{align}
In the last line, we take the limit $\delta\rightarrow 0$. Thus, $O(\delta^{-1})$ and $O(\delta^{0})$ terms are non-zero in Eq.~\eqref{RDM_0;2_1photon_photocurrent}. The careful treatment can be found in the prior studies where the perturbative calculations are performed on the basis of the Bloch states~\cite{Sipe2000,De_Juan2020,Watanabe2021}. $O(\delta^{0})$ terms are given by
        \begin{align}
            2\sigma^{\alpha;\beta\gamma}_\text{intI}
                &=\frac{1}{\Omega^2+  \eta^2}  \sum_{a,c} -\frac{1}{4} \Delta^{\alpha}_{ac}\vj^\beta_{ac}\vj^{\gamma}_{ca}f_{ac} \frac{1}{(\Omega  - i \eta -E_{ac})^2} + \left[ \left( \beta,\gamma,\Omega \right) \leftrightarrow \left( \gamma,\beta,-\Omega \right)  \right],\\
                &=\frac{1}{\Omega^2+  \eta^2}  \sum_{a,c} \frac{1}{4}  \vj^\beta_{ac}\vj^{\gamma}_{ca}f_{ac} \partial_{\alpha} \frac{1}{\Omega  - i \eta -E_{ac}} + \left[ \left( \beta,\gamma,\Omega \right) \leftrightarrow \left( \gamma,\beta,-\Omega \right)  \right],\\
                &=\frac{1}{\Omega^2+  \eta^2}  \sum_{a,c}  \frac{1}{2}  \vj^\beta_{ac}\vj^{\gamma}_{ca}f_{ac} \partial_{\alpha} \text{P} \frac{1}{\Omega -E_{ac}}. \label{injection_reactive_term}
        \end{align}
This contribution will be discussed later.

Taking the $O(\delta^{-1})$ term, we obtain 
    \begin{align}
    2\sigma^{\alpha;\beta\gamma}_\text{inj}
        &=\frac{1/(4i\eta )}{\Omega^2 + \eta^2}  \sum_{a\neq c} -2i\pi \Delta^{\alpha}_{ac} \vj_{ca}^\beta \vj^\gamma_{ac}f_{ac} \delta ( \Omega - E_{ac}),\\
        &=\frac{1/(4i\eta )}{\Omega^2 + \eta^2}  \sum_{a\neq c} -2i\pi \Delta^{\alpha}_{ac} E_{ca}^2 \xi_{ca}^\beta \xi^\gamma_{ac} f_{ac} \delta ( \Omega - E_{ac}),\\
        &=\frac{\Omega^2 /(4i\eta )}{\Omega^2 + \eta^2}  \sum_{a\neq c} -2i\pi \Delta^{\alpha}_{ac}  \xi_{ca}^\beta \xi^\gamma_{ac} f_{ac} \delta ( \Omega - E_{ac}),\\
        &=\frac{-\pi}{2\eta }  \sum_{a\neq c} \Delta^{\alpha}_{ac} \left( g_{ca}^{\beta\gamma} -\frac{i}{2}\Omega_{ca}^{\beta\gamma} \right) f_{ac} \delta ( \Omega - E_{ac}).\label{total_injection_current}
      \end{align} 
We defined the velocity difference matrix $\Delta^\alpha_{ac} = \vj^\alpha_{aa}-\vj^\alpha_{cc} = -\partial_{\alpha} E_{ac}$ and used Eq.~\eqref{hellmann_feynman_velocity_vector} and Eq.~\eqref{quantum_metric_and_Berry_curvature_in_vector_potential} for the band-resolved quantum metric and Berry curvature.
For a simplified notation, we suppress $\lambda$ for the connection, derivative, and geometric quantities.

The obtained component $\sigma_\text{inj}$ is the so-called injection current. We can decompose the injection current into two terms; one is determined by the quantum metric and symmetric under the interchange $\beta \leftrightarrow \gamma$ (photocurrent induced by linearly-polarized light, LP-photocurrent), while the other determined by the Berry curvature is anti-symmetric (photocurrent induced by circularly-polarized light, CP-photocurrent). Labeling these photocurrents by ``$\text{Minj}$'' and ''$\text{Einj}$'', we rewrite the injection current term by 
        \begin{align}
            &\sigma^{\alpha;\beta\gamma}_\text{inj} = \sigma^{\alpha;\beta\gamma}_\text{Minj} +\sigma^{\alpha;\beta\gamma}_\text{Einj}, \label{injection_decomposition}\\
            &2\sigma^{\alpha;\beta\gamma}_\text{Einj}=   \frac{i\pi}{4\eta}  \sum_{a\neq b} \Delta^{\alpha}_{ab}  \Omega_{ba}^{\beta \gamma} f_{ab} \delta (\Omega - E_{ab}), \label{App_electric_injection} \\
            &2\sigma^{\alpha;\beta\gamma}_\text{Minj}=   \frac{-\pi}{2\eta}  \sum_{a\neq b} \Delta^{\alpha}_{ab} g_{ba}^{\beta\gamma} f_{ab} \delta (\Omega - E_{ab}).\label{App_magnetic_injection}
        \end{align}
Because the relevant geometric quantities have the contrasting symmetry property, the LP-(CP-)photocurrent is prohibited by the \T{} (\PT{}) symmetry~\cite{Zhang2019}. Conversely, the LP-(CP-)photocurrent due to the injection current term is allowed in systems preserving the \PT{} (\T{}) symmetry, and it is called magnetic (electric) injection current.
Although the injection current formula shows the diverging behavior due to $\eta = +0$, the scattering effect suppresses this divergence~\cite{DeJuan2016}.
The regularized expression is given by
		\begin{equation}
            2\sigma^{\alpha;\beta\gamma}_\text{inj} =
                \frac{-\pi}{2i \gamma }  \sum_{a\neq b} \Delta^{\alpha}_{ab} \left( g_{ba}^{\beta\gamma} -\frac{i}{2}\Omega_{ba}^{\beta\gamma} \right) f_{ab} \mathcal{L} \left(  \Omega - E_{ab} \right). \label{injection_current_total_regularized}
        \end{equation}
We defined the Lorentzian function parametrized by the scattering rate $\gamma$, 
		\begin{equation}
            \mathcal{L} (x) = \frac{\gamma}{\pi}\frac{1}{\gamma^2 + x^2}.
        \end{equation}
Finally, we obtain the two components from Eq.~\eqref{RDM_0;2_1photon_photocurrent} with the intralevel transition condition $E_a = E_b$ as
        \begin{equation}
            \sigma^{\alpha;\beta\gamma}_{d,{\rm intra}} =\sigma^{\alpha;\beta\gamma}_\text{inj} + \sigma^{\alpha;\beta\gamma}_\text{intI}.
        \end{equation}

Next, we consider the interlevel transition process ($E_{ab}\neq 0$) for $\sigma_d$. In this case, we safely take the limit $\delta \rightarrow 0 $. The expression is given by
        \begin{align}
           2 \sigma^{\alpha;\beta\gamma}_{d,{\rm inter}}
                &=\frac{1}{\Omega^2+  \eta^2}  \Biggl[ \sum_{a,b}' \sum_c \frac{1}{2}\frac{\vj^{\alpha}_{ab}}{2i\eta -E_{ba}} \left(   \frac{\vj^\gamma_{bc}\vj^{\beta}_{ca}f_{ac} }{\Omega  + i \eta -E_{ca} }    -    \frac{\vj^\gamma_{ca}\vj^{\beta}_{bc}f_{cb} }{\Omega  + i \eta -E_{bc} }  \right) \Biggr] \notag \\
                &+ \left[ \left( \beta,\gamma,\Omega \right) \leftrightarrow \left( \gamma,\beta,-\Omega \right)  \right]. \label{threeband_process}
        \end{align}
Here, $\Sigma'$ indicates the summation of $a,b$ with the condition $E_{ab}\neq 0$. Making use of Eq.~\eqref{hellmann_feynman_velocity_vector}, we rewrite the current operator by the Berry connection,
		\begin{equation}
        \frac{\vj^\alpha_{ab}}{E_{ab}}\vj^\gamma_{bc} =  - i \xi^{\alpha}_{ab}\vj^\gamma_{bc} =  \Braket{a | \partial_{\alpha} b }\Braket{b | \vj^{\gamma}  |c } =  -
        \Braket{\partial_{\alpha} a  |  b }\Braket{b | \vj^{\gamma}  | c } =  -\Braket{D_{\alpha} a |  b} \Braket{b | \vj^{\gamma}  | c},
        \end{equation}
where we use the covariant derivative 
            \begin{equation}
            \ket{D_{\alpha} a} = \ket{\partial_{\alpha} a} +i \sum_{d}^{E_{ad}=0} \xi^{\alpha}_{da}\ket{d}. \label{equi_energy_covariant_derivative}
            \end{equation}
The gauge degree of freedom associated with the covariant derivative is defined by the eigenstates having the same energy. The degeneracy does not exist in some cases, and hence the covariant derivative is characterized by the $U(1)$ gauge~\cite{Von_Baltz1981}. On the other hand, for the \PT{} symmetric and spinful systems, the \PT{}-ensured Kramers degeneracy leads to the $U(2)$ covariant derivative~\cite{Watanabe2021} even in the presence of the variational parameter $\bm{\lambda}$. With the covariant derivative, Eq.~\eqref{threeband_process} is transformed as
\begin{align}
    2\sigma^{\alpha;\beta\gamma}_{d,{\rm inter}}
        &=-\frac{1}{2\left( \Omega^2+  \eta^2 \right)}  \Biggl[ \sum_{a,c}  \frac{ \Braket{D_{\alpha} a | \vj^\gamma | c} \vj^{\beta}_{ca}f_{ac} }{\Omega  + i \eta -E_{ca} }    +  \sum_{b,c} \frac{  \Braket{c | \vj^\gamma | D_{\beta} b }\vj^{\beta}_{bc}f_{cb} }{\Omega  + i \eta - E_{bc} }  \Biggr] \notag \\
        &+ \left[ \left( \beta,\gamma,\Omega \right) \leftrightarrow \left( \gamma,\beta,-\Omega \right)  \right],\\
        &=-\frac{1}{2\left( \Omega^2+  \eta^2 \right)}   \sum_{a,b} \Biggl[  \Braket{D_{\alpha} a | \vj^\gamma | b} \vj^{\beta}_{ba} f_{ab} + \Braket{ a | \vj^\gamma |D_{\alpha} b } \vj^{\beta}_{ba}f_{ab} \Biggr] \notag \\
        &~~~~~~\times \left(  \text{P} \frac{1}{\Omega  -E_{ba} } -i\pi \delta ( \Omega -E_{ba})      \right)   + \left[ \left( \beta,\gamma,\Omega \right) \leftrightarrow \left( \gamma,\beta,-\Omega \right)  \right].\label{0;2_1photon_shift}
  \end{align}
Introducing the Berry connection $\mathcal{A}^\alpha$ associated with the covariant derivative $D_\alpha$, we symbolically define the covariant derivative of operators by 
		\begin{equation}
    [D_{\alpha} O]_{ab} = \partial_{\alpha} O_{ab}  - i \left[ \cobc^{\alpha}, O \right]_{ab}.
        \end{equation}
Accordingly, we obtain 
        \begin{equation}
            \Braket{D_{\alpha} a | \vj^\gamma |b}  + \Braket{ a | \vj^\gamma |D_{\alpha} b}  = [D_{\alpha} \vj^\gamma]_{ab} - (\partial_{\alpha} \vj^\gamma)_{ab} = [D_{\alpha} \vj^\gamma]_{ab} + \vj^{\alpha\gamma}_{ab}.\label{equiene_covariant_to_2nd_velocity}
        \end{equation}
Meanwhile, $\sigma_b$ in Eq.~\eqref{RDM_1;1_photocurrent} reads with the limit $\delta\rightarrow 0$,
        \begin{equation}
           2 \sigma_b =  \frac{1}{2\left( \Omega^2+  \eta^2 \right)}    \sum_{a,b} \vj^{\alpha\gamma}_{ab} \vj^\beta_{ba} f_{ab} \left[   \text{P}\frac{1}{\Omega  -E_{ba}} -i\pi \delta ( \Omega -E_{ba}) \right]+ \left[ \left( \beta,\gamma,\Omega \right) \leftrightarrow \left( \gamma,\beta,-\Omega \right)  \right],
        \end{equation}
Thus, by summing this term and Eq.~\eqref{0;2_1photon_shift}, the component including $\vj^{\alpha\gamma}$ in Eq.~\eqref{0;2_1photon_shift} is canceled. The remaining term is given by
		\begin{equation}
            \sigma^{\alpha;\beta\gamma}_{d,inter} + \sigma^{\alpha;\beta\gamma}_{b} = \sigma^{\alpha;\beta\gamma}_\text{abs} +\sigma^{\alpha;\beta\gamma}_\text{rea}, 
        \end{equation}
where 
        \begin{equation}
           2 \sigma^{\alpha;\beta\gamma}_\text{abs} = -\frac{1}{2\left( \Omega^2+  \eta^2 \right)}   \sum_{a,b} -i\pi \Biggl[   [D_{\alpha} \vj^\gamma]_{ab}\vj^\beta_{ba} - [D_{\alpha} \vj^\beta]_{ba}\vj^\gamma_{ab}  \Biggr] f_{ab}  \delta ( \Omega -E_{ba}),
        \end{equation}
and 
        \begin{align}
       2 \sigma^{\alpha;\beta\gamma}_\text{rea} 
            &= -\frac{1}{2\left( \Omega^2+  \eta^2 \right)}   \sum_{a,b}  \Biggl[   [D_{\alpha} \vj^\gamma]_{ab}\vj^\beta_{ba} + [D_{\alpha} \vj^\beta]_{ba}\vj^\gamma_{ab}  \Biggr] f_{ab}  \text{P}\frac{1}{ \Omega  -E_{ba}}.
        \end{align}
$\sigma^{\alpha;\beta\gamma}_\text{abs}$ characterized by the resonant contribution is distinguished from the injection current term [Eq.~\eqref{injection_current_total_regularized}] because it is insensitive to the scattering rate. This term is divided into the LP and CP-photocurrents as
            \begin{align}
            &\sigma^{\alpha;\beta\gamma}_\text{abs} = \sigma^{\alpha;\beta\gamma}_\text{shift} + \sigma^{\alpha;\beta\gamma}_\text{gyro},\\
             &2\sigma^{\alpha;\beta\gamma}_\text{shift} =- \frac{\pi}{2\Omega^2}   \sum_{a\neq b}  \Im{\Biggl[ [D_{\alpha} \vj^\beta]_{ab}\vj^\gamma_{ba} + [D_{\alpha} \vj^\gamma]_{ab}\vj^\beta_{ba}  \Biggr]} f_{ab}  \delta (\Omega -E_{ba}),\label{app_shift_current_velocity}\\
             &2\sigma^{\alpha;\beta\gamma}_\text{gyro} = -\frac{i\pi}{2\Omega^2}   \sum_{a\neq b}  \Re{\Biggl[ [D_{\alpha} \vj^\beta]_{ab}\vj^\gamma_{ba} - [D_{\alpha} \vj^\gamma]_{ab}\vj^\beta_{ba}  \Biggr]} f_{ab}  \delta (\Omega -E_{ba}),\label{app_gyration_current_velocity}
            \end{align}
where we use the identity
    \begin{equation}
        [D_{\alpha} \vj^\gamma]_{ab}\vj^\beta_{ba} - [D_{\alpha} \vj^\beta]_{ba}\vj^\gamma_{ab} = i \Im\,\left(  [D_{\alpha} \vj^\beta]_{ab}\vj^\gamma_{ba} + [D_{\alpha} \vj^\gamma]_{ab}\vj^\beta_{ba} \right) - \Re\,\left(  [D_{\alpha} \vj^\beta]_{ab}\vj^\gamma_{ba} - [D_{\alpha} \vj^\gamma]_{ab}\vj^\beta_{ba} \right).
    \end{equation}
The resulting terms $\sigma_\text{shift}$ and $\sigma_\text{gyro}$ are called shift current~\cite{Von_Baltz1981,Sipe2000,Sturman1992Book} and gyration current~\cite{Watanabe2021,Ahn2020}. Owing to the Hellmann-Feynman relation, the velocity operators are rewritten by the Berry connections as
            \begin{align}
            &2\sigma^{\alpha;\beta\gamma}_\text{shift} = -\frac{\pi}{2}   \sum_{a\neq b}  \Im{\Biggl[ [D_{\alpha} \xi^\beta]_{ab}\xi^\gamma_{ba} + [D_{\alpha} \xi^\gamma]_{ab}\xi^\beta_{ba}  \Biggr]} f_{ab}  \delta (\Omega -E_{ba}),\label{app_shift_current_berry_connection}\\
            &2\sigma^{\alpha;\beta\gamma}_\text{gyro} = -\frac{i\pi}{2}   \sum_{a\neq b}  \Re{\Biggl[ [D_{\alpha} \xi^\beta]_{ab}\xi^\gamma_{ba} - [D_{\alpha} \xi^\gamma]_{ab}\xi^\beta_{ba}  \Biggr]} f_{ab}  \delta (\Omega -E_{ba}).\label{app_gyration_current_berry_connection}
            \end{align}
Following the parallel discussion in Refs.~\cite{Von_Baltz1981,Watanabe2021}, we can show that these photocurrent responses are written by the shift vector and the chiral shift vector, respectively. For the numerical calculations, we approximate the delta function into the Lorentzian function as in Eq.~\eqref{injection_current_total_regularized}.

The off-resonant term $\sigma_\text{rea}$ is added to Eq.~\eqref{injection_reactive_term} and gives 
		\begin{align}
            &2\sigma^{\alpha;\beta\gamma}_\text{intI+rea} =2\sigma^{\alpha;\beta\gamma}_\text{intI} + 2\sigma^{\alpha;\beta\gamma}_\text{rea},\\ 
            &=\frac{1}{2\left( \Omega^2+  \eta^2 \right)} \left[ \sum_{a\neq b} \partial_{\alpha}   \left( \vj^\beta_{ab}\vj^{\gamma}_{ba}  \text{P} \frac{1}{\Omega -E_{ab}} \right)  f_{ab} +  f_{ab}  \text{P}\frac{1}{ \Omega  -E_{ba}}  \left(   i  [\cobc^{\alpha}, \vj^\beta]_{ba}\vj^\gamma_{ab} +i \vj^\beta_{ba} [\cobc^{\alpha}, \vj^\gamma]_{ab}  \right) \right].
        \end{align}
Owing to the definition of the Berry connection $\cobc^\alpha$, the second term vanishes in the last line. Performing the partial integration of the first term, we obtain
		\begin{equation}
            2\sigma^{\alpha;\beta\gamma}_\text{intI+rea}
                = \frac{1}{2\left( \Omega^2+  \eta^2 \right)} \left[   - \sum_{a\neq b}  \vj^\beta_{ab}\vj^{\gamma}_{ba}  \text{P} \frac{1}{\Omega-  E_{ab}}\partial_{\alpha}  f_{ab}  + \partial_{\alpha} \left(  \sum_{a\neq b}  \vj^\beta_{ab}\vj^{\gamma}_{ba}  \text{P} \frac{1}{\Omega -E_{ab}}  f_{ab}  \right) \right].\label{sum_injection_reactive_shift_reactive}
          \end{equation}
We should note that the second term enclosed by the derivative of $\lambda_\alpha$ does not vanish in the BdG Hamiltonian while it is dropped due to the periodicitity in the crystal momentum space in the normal state.
We label the first and second terms in Eq.~\eqref{sum_injection_reactive_shift_reactive} by $\sigma_\text{FS1}$ and $\sigma_\text{SC1}$,
        \begin{align}
            2\sigma^{\alpha;\beta\gamma}_\text{FS1}
                &= - \frac{1}{2\left( \Omega^2+  \eta^2 \right)}  \sum_{a\neq b}  \vj^\beta_{ab}\vj^{\gamma}_{ba}  \text{P} \frac{1}{\Omega-  E_{ab}}\partial_{\alpha}  f_{ab},\\
            2\sigma^{\alpha;\beta\gamma}_\text{SC1}
                &= \frac{1}{2\left( \Omega^2+  \eta^2 \right)}  \partial_{\alpha} \left(  \sum_{a \neq b}  \vj^\beta_{ab}\vj^{\gamma}_{ba}  \text{P} \frac{1}{\Omega -E_{ab}}  f_{ab}  \right).
        \end{align}

Next, we consider $\sigma_a$ [Eq.~\eqref{RDM_2;0_photocurrent}]. Taking $\delta \rightarrow 0$ and using the relation
		\begin{equation}
        \vj^{\alpha\beta\gamma}_{aa} = - \partial_{\alpha} \vj^{\beta\gamma}_{aa}+ i [\xi^{\alpha}, \vj^{\beta\gamma}]_{aa},\label{hellmannn_feynman_3rd_velocity_operator}
        \end{equation}
we obtain 
        \begin{equation}
         2\sigma^{\alpha;\beta\gamma}_{a} = \frac{1}{2(i\Omega - \eta)(-i\Omega - \eta)}  \left[  \sum_a  - \partial_{\alpha} \vj^{\beta\gamma}_{aa}   f_a   +\sum_{a,b} \frac{\vj^\alpha_{ab}\vj^{\beta\gamma}_{ba}f_{ba}}{E_{ab}} \right].
        \end{equation}
The second component in the right hand side is canceled out by $\sigma_c$ [Eq.~\eqref{RDM_0;2_2photon_photocurrent}]. The first component is further transformed by Eq.~\eqref{hellmann_feynman_diamagnetic_current_vector} into 
        \begin{align}
            2\partial_{\alpha} \vj^{\beta\gamma}_{aa} 
                &= -\partial_{\alpha} \partial_{\beta} \vj^{\gamma}_{aa} + i \partial_{\alpha} \left(  [\xi^{\beta}, \vj^{\gamma}]_{aa} \right) - \partial_{\alpha} \partial_{\gamma} \vj^{\beta}_{aa} + i \partial_{\alpha} \left(  [\xi^\gamma, \vj^{\beta}]_{aa} \right),\\
                &= 2 \partial_{\alpha} \partial_{\beta} \partial_{\gamma} \epsilon_{a} + i \partial_{\alpha} \left(  [\xi^{\beta}, \vj^{\gamma}]_{aa} \right)  + i \partial_{\alpha} \left(  [\xi^{\gamma}, \vj^{\beta}]_{aa} \right),\label{2nd_velocity_decomposition}
        \end{align} 
and we accordingly obtain
		\begin{equation}
            2\sigma^{\alpha;\beta\gamma}_{a} + 2\sigma^{\alpha;\beta\gamma}_{c} = -\frac{1}{2\Omega^2} \sum_a f_a \partial_{\alpha} \partial_{\beta} \partial_{\gamma} \epsilon_{a} + \frac{1}{\Omega^2}  \left[  - \sum_{a\neq b} \frac{1}{2} \frac{\vj^\beta_{ab}\vj^\gamma_{ba} }{E_{ab}}\partial_{\alpha} f_{ab} + \partial_{\alpha } \left( \sum_{a\neq b} \frac{1}{2} \frac{\vj^\beta_{ab}\vj^\gamma_{ba} }{E_{ab}}  f_{ab}  \right) \right].\label{drude_other_surface_terms}
        \end{equation}
To see the correspondence between the photocurrent responses in the normal state and those in superconducting state, we perform the partial integration for the first term,
		\begin{equation}
            \frac{1}{2\Omega^2} \left[ \partial_{\beta}   \partial_{\gamma} \left( \sum_a f_a   \partial_{\alpha} \epsilon_{a}  \right) - \partial_{\gamma} \left( \sum_a \partial_{\beta} f_a   \partial_{\alpha} \epsilon_{a} \right)- \partial_{\beta} \left( \sum_a \partial_{\gamma} f_a   \partial_{\alpha} \epsilon_{a} \right) + \sum_a \partial_{\beta}  \partial_{\gamma} f_a  \partial_{\alpha} \epsilon_{a}  \right].
        \end{equation}
What is not differentiated by $\lambda$ is the (nonlinear) Drude term~\cite{Holder2020} denoted by
		\begin{equation}
            2\sigma^{\alpha;\beta\gamma}_\text{D}
            = -\frac{1}{2\Omega^2}  \sum_a \partial_{\alpha} \epsilon_{a} \partial_{\beta} \partial_{\gamma} f_a   .\label{drude_photocurrent}
        \end{equation}
This term can be finite in the normal state. 
The remaining term is finite only in the superconducting state and labeled by
		\begin{equation}
            2\sigma^{\alpha;\beta\gamma}_\text{SC2} =  -\frac{1}{2\Omega^2}  \left[ \partial_{\beta} \partial_{\gamma} \left( \sum_a f_a   \partial_{\alpha} \epsilon_{a}  \right) - \partial_{\gamma} \left( \sum_a \partial_{\beta} f_a   \partial_{\alpha} \epsilon_{a} \right)- \partial_{\beta} \left( \sum_a \partial_{\gamma} f_a   \partial_{\alpha} \epsilon_{a} \right)  \right].
        \end{equation}
We also denote the $\lambda$-derivative term in Eq.~\eqref{drude_other_surface_terms} as
		\begin{equation}
        2\sigma^{\alpha;\beta \gamma}_\text{SC3} = \frac{1}{2\Omega^2} \partial_{\alpha } \left( \sum_{a\neq b} \frac{\vj^\beta_{ab}\vj^\gamma_{ba} }{E_{ab}}  f_{ab}  \right),
        \end{equation}
which is also finite only in the superconducting state. Accordingly, we decompose Eq.~\eqref{drude_other_surface_terms} as
        \begin{equation}
            \sigma^{\alpha;\beta\gamma}_{a} + \sigma^{\alpha;\beta\gamma}_{c}= \sigma^{\alpha;\beta \gamma}_\text{D}+ \sigma^{\alpha;\beta \gamma}_\text{SC2}+ \sigma^{\alpha;\beta \gamma}_\text{SC3}+ \sigma^{\alpha;\beta \gamma}_\text{FS2},
        \end{equation}
where we defined
        \begin{equation}
            2\sigma^{\alpha;\beta \gamma}_\text{FS2} = - \frac{1}{2\Omega^2}  \sum_{a\neq b} \frac{\vj^\beta_{ab}\vj^\gamma_{ba} }{E_{ab}}\partial_{\alpha} f_{ab}.
        \end{equation}
Combining $\sigma_\text{FS1}$ and $\sigma_\text{FS2}$, we obtain a Fermi surface contribution
        \begin{align}
            2\sigma_\text{FS1} +2\sigma_\text{FS2}=
                & - \frac{1}{\Omega}   \sum_{a\neq b} \frac{1}{2} \frac{\vj^\beta_{ab}\vj^\gamma_{ba} }{E_{ab} \left(  \Omega - E_{ab} \right)}\partial_{\alpha} f_{ab},\\
                &= -  \sum_{a\neq b} \frac{1}{2} \frac{\vj^\beta_{ab}\vj^\gamma_{ba} }{E_{ab}^2 } \left( \frac{1}{  \Omega -E_{ab}} - \frac{1}{ \Omega} \right)\partial_{\alpha} f_{ab}.
        \end{align}
In the final line, the first and second components are called intrinsic Fermi surface term~\cite{De_Juan2020,Watanabe2021} and Berry curvature dipole effect~\cite{Moore2010,Sodemann2015}. We denote them as
        \begin{align}
            &\sigma^{\alpha;\beta\gamma}_\text{IFS}= - \frac{1}{2}   \sum_{a\neq b}  \xi^{\beta}_{ab}\xi^{\gamma}_{ba} \frac{1}{  \Omega -E_{ab}} \partial_{\alpha} f_{ab},\\
            &\sigma^{\alpha;\beta\gamma}_\text{BCD}= \frac{1}{2 \Omega}   \sum_{a\neq b}  \xi^{\beta}_{ab}\xi^{\gamma}_{ba}  \partial_{\alpha} f_{ab} = - \frac{i}{2 \Omega}   \sum_{a}  \Omega^{\beta \gamma}_{a}  \partial_{\alpha} f_{a} .\label{berry_curvature_dipole}
        \end{align}
The Berry curvature dipole effect is classified as the CP-photocurrent since it is anti-symmetric under $\beta \leftrightarrow \gamma$. On the other hand, the intrinsic Fermi surface effect can be decomposed into the LP and CP-photocurrents as in the injection current [Eq.~\eqref{injection_decomposition}].
The LP and CP parts are called electric and magnetic intrinsic Fermi surface effects, respectively~\cite{De_Juan2020,Watanabe2021}.

Finally, we obtain the normal photocurrent conductivity
		\begin{equation}
        \sigma_\text{n}^{\alpha;\beta\gamma}
        = \sigma_\text{Einj}^{\alpha;\beta\gamma}+\sigma_\text{Minj}^{\alpha;\beta\gamma}+\sigma_\text{shift}^{\alpha;\beta\gamma}+\sigma_\text{gyro}^{\alpha;\beta\gamma}+\sigma_\text{D}^{\alpha;\beta\gamma}+\sigma_\text{IFS}^{\alpha;\beta\gamma}+\sigma_\text{BCD}^{\alpha;\beta\gamma}.
        \end{equation}
Note that we can easily reproduce the photocurrent response in the normal state by replacing $\bm{\lambda}$ with $\bk$ in the normal photocurrent.
The nonlinear Drude term, intrinsic Fermi surface term, and Berry curvature dipole effect include the derivative of the Fermi distribution function and thus vanish in a gapped system at low temperature.

Summing the contributions enclosed by the $\bla$-differentiation, we obtain the anomalous photocurrent arising from the superconductivity
\begin{align}
    2\sigma_\text{a}^{\alpha;\beta\gamma}
        &= 2\sigma_\text{SC1}^{\alpha;\beta\gamma}+2\sigma_\text{SC2}^{\alpha;\beta\gamma}+2\sigma_\text{SC3}^{\alpha;\beta\gamma},\\
        &=  \frac{1}{2\Omega^2} \partial_{\alpha} \left(  \sum_{a,b}  \vj^\beta_{ab}\vj^{\gamma}_{ba}  \text{P} \frac{1}{\Omega -E_{ab}}  f_{ab}  \right) + \frac{1}{\Omega^2 }\partial_{\alpha } \left( \sum_{a\neq b} \frac{1}{2} \frac{\vj^\beta_{ab}\vj^\gamma_{ba} }{E_{ab}}  f_{ab}  \right) \notag \\
        & - \frac{1}{2\Omega^2}  \left[\partial_{\beta} \partial_{\gamma} \left( \sum_a f (\epsilon_{a})   \partial_{\alpha} \epsilon_{a}  \right) - \partial_{\beta}\left( \sum_a \partial_{\gamma} f (\epsilon_{a})   \partial_{\alpha} \epsilon_{a} \right) - \partial_{\gamma}\left( \sum_a \partial_{\beta} f (\epsilon_{a})   \partial_{\alpha} \epsilon_{a} \right)\right] ,\label{eq2} \\
        &=  \frac{1}{2\Omega^2} \partial_{\alpha} \left[  \sum_{a \neq b}  \vj^\beta_{ab}\vj^{\gamma}_{ba} \left(  \text{P} \frac{1}{\Omega - E_{ab}} + \frac{1}{E_{ab}}  \right)   f_{ab}  \right]  - \frac{1}{\Omega^2}  \partial_{\alpha}  \partial_{\beta} \partial_{\gamma} F_{\bm{\lambda}}\notag \\
        &+ \frac{1}{2\Omega^2}  \left[  \partial_{\beta}\left( \sum_a \partial_{\gamma} f (\epsilon_{a})   \partial_{\alpha} \epsilon_{a} \right) + \partial_{\gamma}\left( \sum_a \partial_{\beta} f (\epsilon_{a})   \partial_{\alpha} \epsilon_{a} \right)\right] .\label{photoresponse_superconductor}
    \end{align}
This is the new contribution to the photocurrent clarified by this work. Assuming the gapped system at low temperature, we here ignore the Fermi surface contribution in the second line of Eq.~\eqref{photoresponse_superconductor}. We denote the remaining new contributions as
		\begin{align}
        &2\sigma_\text{NRSF}^{\alpha;\beta\gamma} = -\frac{1}{\Omega^2}  \partial_{\alpha}  \partial_{\beta} \partial_{\gamma} F_{\bm{\lambda}},\label{app_nonreciprocal_SF}\\
        &2\sigma_\text{CD}^{\alpha;\beta\gamma} = \frac{1}{2\Omega^2} \partial_{\alpha} \left[  \sum_{a\neq b}  \vj^\beta_{ab}\vj^{\gamma}_{ba} \left(  \text{P} \frac{1}{\Omega - E_{ab}} + \frac{1}{E_{ab}}  \right)   f_{ab}  \right],\label{app_conductivity_derivative}
        \end{align}
where ``NRSF`` denotes nonreciprocal superfluid density while ``CD'' means conductivity derivative.
The nonreciprocal superfluid density is defined as $f^{\alpha\beta\gamma} = \limvec \partial_{\alpha}  \partial_{\beta} \partial_{\gamma} F_{\bm{\lambda}}$.

It is convenient for numerical computations to rewrite the formulas by generalized velocity operators.
A straightforward example is the injection current contributions in Eq.~\eqref{total_injection_current}.
Although Eq.~\eqref{total_injection_current} implicitly includes the infinitesimal variational parameter $\bm{\lambda}$, we can obtain the formula without $\bm{\lambda}$ by taking the limit $\bm{\lambda} \rightarrow \bm{0}$.
The resulting formula is written by the generalized velocity operators and energy spectrum of the BdG Hamiltonian.
The quantum metric and Berry curvature are given by
		\begin{align}
            &g^{\alpha \beta}_{ab} = \frac{1}{E_{ab}^2} \Re{[\vj^{\alpha}_{ab}\vj^{\beta}_{ba}]},\\
            &\Omega^{\alpha \beta}_{ab} = - \frac{2}{E_{ab}^2} \Im{[\vj^{\alpha}_{ab}\vj^{\beta}_{ba}]},
        \end{align} 
where we use Eq.~\eqref{hellmannn_feryman_velocity} with $a\neq b$.
Since the BdG Hamiltonian can be labeled by the crystal momentum $\bk$, the injection current formula is obtained as
		\begin{equation}
            2\sigma^{\alpha;\beta\gamma}_\text{inj}
            =\frac{-\pi }{2\eta } \sum_\bk \sum_{a,b} \left[ \vj^{\alpha}_{aa} (\bk) - \vj^{\alpha}_{bb}(\bk) \right] \frac{\vj^\beta_{ba}(\bk)\vj^\gamma_{ab}(\bk) }{E_{ba}^2} f_{ab} \delta ( \Omega - E_{ab}),\label{total_injection_current_velocity_representation}
        \end{equation} 
where $a,b$ denote the band indices of Bogoliubov quasiparticles.
Following the parallel discussion, we can derive the numerically convenient formulas for other contributions by transforming the $\bla$-parametrized formulas into what is given by the generalized velocity operators and energy spectrum.
In the case of the shift current and gyration current contributions [Eqs.~\eqref{app_shift_current_berry_connection} and~\eqref{app_gyration_current_berry_connection}], the covariant derivative of the Berry connection is rewritten by
    \begin{equation}
    \left[ D_\alpha  \xi^{\beta} \right]_{ab} 
        = i \vj^\beta_{ab}  \frac{\Delta_{ab}^\alpha }{E_{ab}^2} +  \frac{i}{E_{ab}} \left(   -\vj_{ab}^{\alpha\beta} - \sum_{a\neq c} \frac{\vj^\alpha_{ac} \vj^\beta_{cb}}{E_{ac}}+\sum_{b\neq c} \frac{\vj^\alpha_{cb} \vj^\beta_{ac}}{E_{cb}} \right).
    \end{equation}
  After taking the limit $\bla \rightarrow \bm{0}$, the right hand side becomes a numerically convenient form.

Next, we derive the anomalous photocurrent formulas in the current operator representation. For instance, we take the nonreciprocal superfluid density effect [Eq.~\eqref{app_nonreciprocal_SF}].
When we assume the gapped and low-temperature state for simplicity, we can neglect the term including the derivative of the Fermi distribution function.
The obtained formula for the nonreciprocal superfluid density is 
		\begin{align}
            -2\partial_{\alpha}  \partial_{\beta} \partial_{\gamma} F_{\bm{\lambda}} 
            &=\sum_{a}\vj^{\alpha\beta\gamma}_{aa}f_a + \sum_{a\neq b}\frac{1}{E_{ab}} \left( \vj^{\alpha}_{ab}\vj^{\beta\gamma}_{ba}  + \vj^{\beta}_{ab}\vj^{\gamma\alpha}_{ba} + \vj^{\gamma}_{ab}\vj^{\alpha\beta}_{ba} + c.c. \right)  f_a \notag \\
            &- \sum_{a\neq b}\frac{\Delta^{\gamma}_{ab}}{E_{ab}^2} \left( \vj^{\beta}_{ab}\vj^{\alpha}_{ba} + \vj^{\beta}_{ba}\vj^{\alpha}_{ab} \right) f_a \notag \\
            & + \sum_{a\neq b,c} \frac{1}{E_{ab}E_{ac}} f_a \left[  \vj^\alpha_{ab}\vj^\beta_{bc}\vj^\gamma_{ca} +\vj^\beta_{ab}\vj^\alpha_{bc}\vj^\gamma_{ca} +  c.c \right] \notag \\
            &+ \sum_{b\neq a,c} \frac{1}{E_{ab}E_{bc}} f_a \left[  \vj^\beta_{ab}\vj^\gamma_{bc}\vj^\alpha_{ca} + \vj^\alpha_{ab}\vj^\gamma_{bc}\vj^\beta_{ca} +  c.c \right].\label{NSF_numerical_formula}
        \end{align}
Similarly, we can rewrite the formula for the conductivity derivative effect [Eq.~\eqref{app_conductivity_derivative}] as  
\begin{align}
    &\partial_{\alpha} \left[   \vj^\beta_{ab}\vj^{\gamma}_{ba} \left(  \text{P} \frac{1}{\Omega -E_{ab}} + \frac{1}{E_{ab}}  \right)   f_{ab}  \right]\\
      &=  - \vj^\beta_{ab}\vj^{\gamma}_{ba} \Delta^\alpha_{ab} \left(  \text{P} \frac{1}{\left( \Omega -E_{ab} \right)^2} - \frac{1}{E_{ab}^2}  \right)  f_{ab}\notag \\
F      &+ \left(-\vj^{\alpha\beta}_{ab} \vj^\gamma_{ba} -\vj^{\beta}_{ab} \vj^{\alpha\gamma}_{ba}    \right)  \left(  \text{P} \frac{1}{\Omega -E_{ab}} + \frac{1}{E_{ab}}  \right) f_{ab}\notag \\
      &+ \left[ \sum_{c\neq a}  \left(- \frac{\vj^\alpha_{ac}\vj^\beta_{cb}\vj^\gamma_{ba} }{E_{ac}} +  \frac{\vj^\beta_{ab}\vj^\alpha_{ca}\vj^\gamma_{bc} }{E_{ca}} \right) + \sum_{c\neq b}  \left( \frac{\vj^\alpha_{cb}\vj^\beta_{ac}\vj^\gamma_{ba} }{E_{cb}} -  \frac{\vj^\beta_{ab}\vj^\alpha_{bc}\vj^\gamma_{ca} }{E_{bc}} \right) \right]  \left(  \text{P} \frac{1}{\Omega -E_{ab}} + \frac{1}{E_{ab}}  \right) f_{ab},
    \end{align}
where we omit the prefactor $1/(2\Omega^2)$ and neglect the derivative of the Fermi distribution function.

Note that there is an arbitrariness in decomposing the total photocurrent into the normal and anomalous contributions within the independent particle approximation. For instance, although the Berry curvature dipole effect in Eq.~\eqref{berry_curvature_dipole} is seemingly different from what has been derived in prior works~\cite{Moore2010,Deyo2009,Sodemann2015}, we can reproduce the known expression. Since the Berry curvature is given by the derivative of the intraband Berry connection ($\Omega_a^{\alpha\beta} = \partial_\alpha \xi^\beta_{aa}- \partial_\beta \xi^\alpha_{aa}$), Eq.~\eqref{berry_curvature_dipole} is rewritten by
    \begin{align}
    2\sigma^{\alpha;\beta\gamma}_\text{BCD}
        &= - i \frac{1}{2 \Omega}  \left( \sum_{a}  \partial_{\alpha} \left( \Omega^{\beta \gamma}_{a} f_{a} \right)  - \partial_{\alpha}  \Omega^{\beta \gamma}_{a} f_{a}\right),\\
        &= - i \frac{1}{2 \Omega}   \sum_{a}  \left( \partial_{\alpha} \left( \Omega^{\beta \gamma}_{a} f_{a} \right)  - \left(  \partial_{\beta }\Omega_a^{\alpha \gamma} - \partial_{\gamma }\Omega_a^{\alpha \beta} \right) f_{a}\right),\\
        &= - i \frac{1}{2 \Omega}   \sum_{a}  \left[ \partial_{\alpha} \left( \Omega^{\beta \gamma}_{a} f_{a} \right)  -  \partial_{\beta } \left( \Omega_a^{\alpha \gamma}f_a \right) + \partial_{\gamma }\left( \Omega_a^{\alpha \beta} f_{a}\right) + \left(  \Omega_a^{\alpha \gamma} \partial_{\beta } f_a - \Omega_a^{\alpha \beta}\partial_{\gamma } f_{a} \right) \right],
  \end{align}
where we obtain the ``conventional'' Berry curvature dipole contribution 
		\begin{equation}
        2\sigma^{\alpha;\beta\gamma}_\text{cBCD} =  i \frac{1}{2 \Omega}\sum_a  \left( \Omega_a^{\alpha\beta} \partial_{\gamma} f_a - \Omega_a^{\alpha \gamma}\partial_{\beta } f_a \right), \label{berry_curvature_dipole_conventional}
        \end{equation}
The remaining contributions are enclosed by the $\lambda$-derivative,
		\begin{equation}
            - i \frac{1}{2 \Omega}   \sum_{a} \left[ \partial_{\alpha} \left( \Omega^{\beta \gamma}_{a} f_{a} \right)  -  \partial_{\beta } \left( \Omega_a^{\alpha \gamma}f_a \right) + \partial_{\gamma }\left( \Omega_a^{\alpha \beta} f_{a}\right)\right],\label{anomalous_berry_curvature_dipole_effect}
        \end{equation}
which should vanish in the normal state. Accordingly, the two formulas for the Berry curvature dipole effect [Eqs.~\eqref{berry_curvature_dipole} and~\eqref{berry_curvature_dipole_conventional}] are equivalent in the normal state. On the other hand, when we adopt the formula in Eq.~\eqref{berry_curvature_dipole_conventional} as the Berry curvature dipole effect, it is necessary to include Eq.~\eqref{anomalous_berry_curvature_dipole_effect} in the anomalous contribution. Such arbitrariness in the normal and anomalous photocurrents may occur because we formulate the response function without any scattering effect. This difficulty is resolved by the Green function method in Appendix~\ref{App_Sec_Green_function_method_derivation}.

Some of the normal photocurrents diverge in the static limit, for instance, the Drude photocurrent in Eq.~\eqref{drude_photocurrent}. On the other hand, when the scattering effect is properly introduced, the normal photocurrent is bounded to be finite even in the zero-frequency limit ($\Omega \rightarrow 0$)~\cite{Michishita2020}. On the contrary, the anomalous photocurrent such as the nonreciprocal superfluid density effect should diverge in the low-frequency limit even with scattering effects since the divergence arises from rewriting the vector potential by the electric field as $\bm{E} = i \Omega \bm{A}$.

\section{Derivation by Green function method}
\label{App_Sec_Green_function_method_derivation}

We formulate the nonreciprocal optical response in superconductors based on the Green function method.
The Green function formalism for the nonlinear conductivity has been reported in prior works~\cite{Michishita2020,Holder2020,Parker2019}, and we apply it to the BdG Hamiltonian.
Starting from the Hamiltonian having the electron correlation, we take the molecular field approximation for the Cooper channel of the electron-electron interaction.
Accordingly, the action is given by
		\begin{equation}
                S = \int d\tau \int d\bm{r}\, \frac{1}{2} \,\overline{\bm{\Psi}} \left( \partial_\tau + H_\text{BdG}  \right) \bm{\Psi},
                \end{equation}
where we suppressed a constant arising from the molecular field approximation and neglected other channels. $\bm{\Psi}$ is the Nambu Grassman vector consisting of the Grassmann numbers for electrons and holes, $\bm{\Psi}^T = \left(  \psi_1,\cdots \psi_n,\overline{\psi}_1,\cdots ,\overline{\psi}_n \right)$.
Since the action is quadratic in the Grassmann numbers, we can analytically perform the Grassmann integration. The partition function is obtained as
		\begin{equation}
                Z = \Tr{e^{-\beta \mathcal{H}}} = \int \mathcal{D}\overline{\psi}\mathcal{D}{\psi} e^{-S} = \int d\tau \int d\bm{r} \frac{1}{2} \Tr{\log{\mathcal{G}^{-1} (\tau)}}.
                \end{equation}
We introduced the Gor'kov Green function $\mathcal{G}$
        \begin{equation}
        \mathcal{G}^{-1} = 
                  \begin{pmatrix}
                  \partial_\tau + H_\text{N} & \Delta \\
                  \Delta^\dagger & \partial_\tau -\left( H_\text{N} \right)^T
                  \end{pmatrix},
          \end{equation}
where the imaginary-time derivative is performed on the right hand side. It is necessary for calculations of the conductivity to take into account the electromagnetic fields. In this work, we introduce the minimal coupling between the vector potential and the canonical momentum in the normal-state Hamiltonian as 
		\begin{equation}
                H_\text{N} [\hat{\bm{p}}] \rightarrow   H_\text{N} [\hat{\bm{p}}-\bA (\tau)] 
                \equiv H_\text{N} (\bA(\tau)).
                \end{equation}
We take the vector potential which yields the photo-electric field in the real-time domain as $\bm{E}= -\partial_t \bm{A} (t)$. Correspondingly, the Gor'kov Green function is replaced with $\mathcal{G}_\bA$ whose the normal-state Hamiltonian is $H_\text{N} (\bA(\tau))$. 

The expectation value of the electric current is obtained in a straightforward way once the partition function is given. The electric current is calculated as 
		\begin{align}
                \Braket{ \mathcal{J}^\alpha (\tau)}
                &=- \frac{\delta}{\delta A_\alpha (\tau)}\int d\tau' \int d \bm{r}' \frac{1}{2} \Tr{[\log{\mathcal{G}_\bA^{-1}}(\tau') ]},\\
                &=\frac{1}{2} \int d \bm{r} \Tr{ \vj^\alpha_{\bA} (\tau) \mathcal{G}_\bA (\tau) },\label{App_electric_current_SC_expression}
                \end{align}
where the coefficient $1/2$ is due to the particle hole doubling in the BdG Hamiltonian. $\vj^\alpha_{\bA}$ is the current operator for the BdG Hamiltonian and defined as
                \begin{equation}
                \vj^\alpha_\bA(\tau) =- \frac{\delta}{\delta A_\alpha (\tau)}
                \begin{pmatrix}
                        H_\text{N} (\bA (\tau))  & 0 \\
                        0 & -\left[ H_\text{N}(\bA(\tau))\right]^T
                \end{pmatrix}.
                \end{equation}
The linear and nonlinear conductivities are systematically calculated by expanding the current operator $\vj^\alpha_{\bA}$ and Green function $ \mathcal{G}_\bA$ with respect to the vector potential $\bA$. The derivation is similarly performed as in Ref.~\cite{Michishita2020}.
We show the final result of the NRO conductivity
        \begin{align}
        &2\sigma^{\alpha ;\beta \gamma}\left(\omega ; \omega_{1}, \omega_{2}\right)\notag \\ 
            &=\frac{1}{\omega_{1} \omega_{2}} \int_{-\infty}^{\infty} \frac{d \Omega}{2 \pi i} f(\Omega)  \frac{1}{2}\left\{\frac{1}{2} \Tr \left[\vj^{\alpha \beta \gamma}\left(G^R(\Omega)-G^A(\Omega)\right)\right]\right. \\
            &+\Tr \left[\vj^{\alpha \beta} G^R\left(\Omega+\omega_{2}\right) \vj^{\gamma}\left(G^R(\Omega)-G^A(\Omega)\right)+\vj^{\alpha \beta}\left(G^R(\Omega)-G^A(\Omega)\right) \vj^{\gamma} G^A\left(\Omega-\omega_{2}\right)\right] \\
            &+\frac{1}{2}\Tr\left[\vj^{\alpha} G^R \left(\Omega+\omega \right) \vj^{\beta \gamma}\left(G^R(\Omega)-G^A(\Omega)\right)+\vj^{\alpha}\left(G^R(\Omega)-G^A(\Omega)\right) \vj^{\beta \gamma} G^A\left(\Omega-\omega \right)\right] \\
            &+\Tr\left[\vj^{\alpha} G^R\left(\Omega+\omega \right) \vj^{\beta} G^R\left(\Omega+\omega_{2}\right) \vj^{\gamma}\left(G^R(\Omega)-G^A(\Omega)\right)\right. \\
            &+ \vj^{\alpha} G^R\left(\Omega+\omega_{1}\right) \vj^{\beta}\left(G^R(\Omega)-G^A(\Omega)\right) \vj^{\gamma} G^A\left(\Omega-\omega_{2}\right) \\
            &+\left.\vj^{\alpha}\left(G^R(\Omega)-G^A(\Omega)\right) \vj^{\beta} G^A\left(\Omega-\omega_{1}\right) \vj^{\gamma} G^A\left(\Omega-\omega \right)\right] \\
            &\left.+\left[\left(\beta, \omega_{1}\right) \leftrightarrow\left(\gamma, \omega_{2}\right)\right]\right\},\notag
        \end{align}
where $\Tr{}$ represents the trace over eigenstates of the Bogoliubov quasiparticle. We introduced the retarded ($G^R$) and advanced ($G^A$) Green functions as
		\begin{equation}
                G^{R} (\omega) = \frac{1}{\omega +  i\eta - H_\text{BdG}},~~~G^A (\omega) = \left( G^R (\omega) \right)^\dagger.
                \end{equation}
Here, $\eta>0$ is an infinitesimal parameter building the causality into the Green functions, $H_\text{BdG}$ is the BdG Hamiltonian having no electromagnetic perturbation,
and $\left[\left(\beta, \omega_{1}\right) \leftrightarrow\left(\gamma, \omega_{2}\right)\right]$ means the symmetrization of the applied electric fields due to the intrinsic permutation symmetry of external fields. The doubling of the conductivity ($2\sigma^{\alpha;\beta\gamma}$) respects the coefficient $1/2$ in Eq.~\eqref{App_electric_current_SC_expression}.  

The low-frequency behavior of the NRO conductivity is of current interest.
Thus, we take the $(\omega,\omega_1,\omega_2)$- expansion from the static limit with the frequency conservation condition, $\omega = \omega_1 +\omega_2$.
The leading terms are given by
		\begin{equation}
                       2 \sigma^{\alpha ;\beta \gamma} \left(\omega_1+\omega_2 ; \omega_{1}, \omega_{2}\right)  = \frac{1}{2\omega_1\omega_2} A_{\alpha\beta\gamma} + \frac{1}{2\omega_2}B_{\alpha\beta\gamma}+ \frac{1}{2\omega_1}B'_{\alpha\beta\gamma} + O(\omega^0).\label{App_second_cond_divergent}
                \end{equation}
The coefficients are obtained as
\begin{align}
        A_{\alpha\beta\gamma} 
            &= \int_{-\infty}^{\infty} \frac{d \Omega}{2 \pi i} f(\Omega) \left\{\frac{1}{2} \Tr \left[\vj^{\alpha \beta \gamma}\left(G^R(\Omega)-G^A(\Omega)\right)\right]\right.\label{App_NRSF_third_velocity_term} \\
            &+\Tr \left[\vj^{\alpha \beta} G^R\left(\Omega\right) \vj^{\gamma}\left(G^R(\Omega)-G^A(\Omega)\right)+\vj^{\alpha \beta}\left(G^R(\Omega)-G^A(\Omega)\right) \vj^{\gamma} G^A\left(\Omega\right)\right] \\
            &+\frac{1}{2}\Tr\left[\vj^{\alpha} G^R \left(\Omega+\omega \right) \vj^{\beta \gamma}\left(G^R(\Omega)-G^A(\Omega)\right)+\vj^{\alpha}\left(G^R(\Omega)-G^A(\Omega)\right) \vj^{\beta \gamma} G^A\left(\Omega-\omega \right)\right] \label{App_NRSF_para_to_dens_dia_term} \\
            &+\Tr\left[\vj^{\alpha} G^R\left(\Omega \right) \vj^{\beta} G^R\left(\Omega\right) \vj^{\gamma}\left(G^R(\Omega)-G^A(\Omega)\right)\right. \\
            &+ \vj^{\alpha} G^R\left(\Omega\right) \vj^{\beta}\left(G^R(\Omega)-G^A(\Omega)\right) \vj^{\gamma} G^A\left(\Omega\right) \\
            &+\left.\vj^{\alpha}\left(G^R(\Omega)-G^A(\Omega)\right) \vj^{\beta} G^A\left(\Omega \right) \vj^{\gamma} G^A\left(\Omega \right)\right] \\
            &\left.+\left( \beta  \leftrightarrow  \gamma \right) \right\}.\notag
        \end{align}
and 
\begin{align}
        B_{\alpha\beta\gamma} =
           &\int_{-\infty}^{\infty} \frac{d \Omega}{2 \pi i} f(\Omega)   \Bigl\{ \Tr \left[\vj^{\alpha \beta} F^{R} (\Omega) \vj^{\gamma} \left(G^{R}(\Omega)-G^{A}(\Omega)\right) - \vj^{\alpha \beta}\left(G^{R}(\Omega)-G^{A}(\Omega)\right) \vj^{\gamma} F^{A}\left(\Omega \right)\right]\label{App_cond_deriv_one_term} \\
           &
           + \Tr \left[ \vj^{\alpha} F^{R}\left(\Omega \right) \vj^{\beta \gamma} \left(G^{R}(\Omega)-G^{A}(\Omega)\right) - \vj^{\alpha} \left(G^{R}(\Omega)-G^{A}(\Omega)\right) \vj^{\beta \gamma} F^{A}\left(\Omega \right)\right] \\
            &
            +\Tr \left[
                  \vj^{\alpha} F^{R}\left(\Omega \right) \vj^{\beta} G^{R}\left(\Omega\right) \vj^{\gamma}\left(G^{R}(\Omega)-G^{A}(\Omega)\right)
                  + \vj^{\alpha} G^{R}\left(\Omega \right) \vj^{\beta} F^{R}\left(\Omega\right) \vj^{\gamma}\left(G^{R}(\Omega)-G^{A}(\Omega)\right)\right. \notag \\
            &~~~~~\left.
                  + \vj^{\alpha} F^{R}\left(\Omega \right) \vj^{\gamma} G^{R}\left(\Omega\right) \vj^{\beta}\left(G^{R}(\Omega)-G^{A}(\Omega)\right)\right. \notag \\
            &~~~~~\left.
                  - \vj^{\alpha} G^{R}\left(\Omega \right) \vj^{\beta}\left(G^{R}(\Omega)-G^{A}(\Omega)\right) \vj^{\gamma} F^{A}\left(\Omega\right)
                  + \vj^{\alpha} F^{R}\left(\Omega \right) \vj^{\beta}\left(G^{R}(\Omega)-G^{A}(\Omega)\right) \vj^{\gamma} G^{A}\left(\Omega\right) \right.\notag \\
            &~~~~~\left.
                  - \vj^{\alpha}\left(G^{R}(\Omega)-G^{A}(\Omega)\right) \vj^{\beta} G^{A}\left(\Omega \right) \vj^{\gamma} F^{A}\left(\Omega \right) 
                  - \vj^{\alpha}\left(G^{R}(\Omega)-G^{A}(\Omega)\right) \vj^{\gamma} G^{A}\left(\Omega \right) \vj^{\beta} F^{A}\left(\Omega \right)\right. \notag \\
            &~~~~~\left.
                  - \vj^{\alpha}\left(G^{R}(\Omega)-G^{A}(\Omega)\right) \vj^{\gamma} F^{A}\left(\Omega \right) \vj^{\beta} G^{A}\left(\Omega \right) \right] \Bigr\}.
       \end{align}
$F^{R(A)}(\omega) = \partial_{\omega} G^{R(A)}$ is the frequency derivative of the Green function. The coefficient $B'_{\alpha\beta\gamma}$ is obtained by the permutation $\beta \leftrightarrow \gamma$ in $B_{\alpha\beta\gamma}$. 

Simplification of these coefficients is performed by making use of the variational parameter $\bla$ as in the density matrix formulation.
Let us consider the adiabatic switching of $\bla$ in the BdG Hamiltonian as 
		\begin{equation}
                H_\text{BdG} = \limvec H_\text{BdG}^{(\bla)}.
                \end{equation} 
The right hand side is written by
		\begin{equation}
                H_\text{BdG}^{(\bla)}=
                \begin{pmatrix}
                        H_\text{N}^{\bk-\bla} & \hat{\Delta}_\bk\\
                        \hat{\Delta}_\bk^{\dag} & -(H_\text{N}^{-\bk-\bla})^T
                \end{pmatrix},
                \end{equation}
in the crystal momentum ($\bk$) basis. Normal-state Hamiltonian and pair potentials are accordingly given in the $\bk$ representation. Following the definition of the current operators, we can rewrite the operators by
        \begin{align}
        &J^\alpha =\limvec J^\alpha_\bla= \limvec -\frac{\partial H_\text{BdG}^{(\bla)}}{\partial \lambda_\alpha },\\
        &J^{\alpha\beta}=\limvec J^{\alpha\beta}_\bla = \limvec \frac{\partial^2 H_\text{BdG}^{(\bla)}}{\partial \lambda_\alpha \partial \lambda_\beta},\\
        &J^{\alpha\beta\gamma}=\limvec J^{\alpha\beta\gamma}_\bla =\limvec -\frac{\partial^3 H_\text{BdG}^{(\bla)}}{\partial \lambda_\alpha \partial \lambda_\beta \partial \lambda_\gamma}.
        \end{align}
Owing to the correction to the BdG Hamiltonian, the Green functions are modified. For instance, the retarded Green function is 
		\begin{equation}
                G^R_\bla (\omega) = \frac{1}{\omega+i\eta -H_\text{BdG}^{(\bla)}}.
                \end{equation}
Differentiating the Green functions by $\bla$, we obtain
		\begin{equation}
                \frac{\partial G^R_\bla  (\omega)}{\partial \lambda_\alpha} = -G^R_\bla  (\omega) \vj^{\alpha}_\bla G^R_\bla (\omega).\label{App_Green_fuction_A_deriv}
                \end{equation}

When we make use of the $\bla$-switching and the relation \eqref{App_Green_fuction_A_deriv}, we can simplify the coefficients $A_{\alpha\beta\gamma},~B_{\alpha\beta\gamma},$ and $B'_{\alpha\beta\gamma}$. First, we consider $A_{\alpha\beta\gamma}$. Based on the adiabatic switching of $\bla$, the current operators and Green functions are converted into those labeled by $\bla$. For instance, we take the line \eqref{App_NRSF_third_velocity_term}. The third-order current operator is related with the diamagnetic current operator as $\vj^{\alpha\beta \gamma}_\bla = - \partial_{\lambda_\alpha} \vj^{\beta\gamma}_\bla$. Accordingly,
		\begin{align}
                &\Tr \left[\vj^{\alpha \beta \gamma}_{\bla}\left(G^R_\bla(\Omega)-G^A_\bla(\Omega)\right)\right]\notag \\
                &=- \partial_{\lambda_\alpha} \left(  \Tr \left[\vj^{\beta \gamma}_{\bla}\left(G^R_\bla(\Omega)-G^A_\bla(\Omega)\right)\right] \right) + \Tr \left[\vj^{\beta \gamma}_{\bla} \partial_{\lambda_\alpha} \left(G^R_\bla(\Omega)-G^A_\bla(\Omega)\right)\right],\\ 
                        &=- \partial_{\lambda_\alpha} \left(  \Tr \left[\vj^{\beta \gamma}_{\bla}\left(G^R_\bla(\Omega)-G^A_\bla(\Omega)\right)\right] \right) - \Tr \left[\vj^{\beta \gamma}_{\bla} \left(G^R_\bla(\Omega) \vj^\alpha_{\bla} G^R_\bla(\Omega)-G^A_\bla(\Omega)\vj^\alpha_{\bla} G^A_\bla(\Omega)\right)\right],
                \end{align}
Using the cyclic property of trace, the second term in the last line is canceled with Eq.~\eqref{App_NRSF_para_to_dens_dia_term}.
Performing a similar algebra, we finally rewrite $A_{\alpha\beta\gamma}$ by
		\begin{align}
                A_{\alpha\beta\gamma} 
                        &=\int \frac{d\Omega}{2 \pi i} f (\Omega) \limvec \partial_{\lambda_\beta}\partial_{\lambda_\gamma} \Tr{\left[ \vj^\alpha_{\bla} \left( G ^R_\bla (\Omega) - G^A_\bla (\Omega) \right) \right]},\\
                        &=-  \limvec \partial_{\lambda_\beta}\partial_{\lambda_\gamma} \Tr{\left[ \vj^\alpha_{\bla} f (H_\text{BdG}^{(\bla)}) \right]},\\
                        &= \limvec \partial_{\lambda_\alpha} \partial_{\lambda_\beta}\partial_{\lambda_\gamma}  \left(-\frac{1}{\beta} \Tr{\left[ \log{\left( 1+e^{-\beta H_\text{BdG}^{(\bla)}} \right)} \right]} \right),\\
                        &= 2 \limvec \partial_{\lambda_\alpha} \partial_{\lambda_\beta}\partial_{\lambda_\gamma}  F_\bla.
                \end{align}
As a result, we reproduce the formula for the nonreciprocal superfluid density effect given by
		\begin{equation}
            2 \sigma^{\alpha ;\beta \gamma}_\text{NRSF} \left(\omega_1+\omega_2 ; \omega_{1}, \omega_{2}\right)=  \frac{1}{2\omega_1 \omega_2} A_{\alpha\beta\gamma} =  \frac{1}{\omega_1 \omega_2} f^{\alpha\beta\gamma}.\label{Green_function_NRSF_effect}
                \end{equation}
The obtained formula is the generalized expression of Eq.~\eqref{app_nonreciprocal_SF} and captures the nonreciprocal superfluid density effect in general NRO responses including those other than the photocurrent generation $\omega_1 =-\omega_2$. 

Next, we consider the coefficients of the $O(\omega^{-1})$ term in Eq.~\eqref{App_second_cond_divergent}. Similarly to $A_{\alpha\beta\gamma}$, the coefficient $B_{\alpha\beta\gamma}$ is simplified as
		\begin{equation}
                B_{\alpha\beta \gamma} =   \int \frac{d\Omega}{2 \pi i} f (\Omega) \limvec \partial_{\lambda_\beta} \Tr{\left[ \vj^\alpha_{\bla}  \left( G_\bla^R (\Omega) - G_\bla^A (\Omega) \right) \vj^\gamma_{\bla} F_\bla^A  -\vj^\alpha_{\bla} F_\bla^R  \vj^\gamma_{\bla} \left( G_\bla^R (\Omega) - G_\bla^A (\Omega) \right)  \right]}.\label{App_B_conductivity_derivative_bare}
                \end{equation}
Since the $\lambda_\beta$-derivative is taken over the whole expression, this contribution vanishes in the normal state.
Interestingly, this formula has the similar form to the regular part of the linear static conductivity. The expression is given by
		\begin{equation}
                \sigma_{\alpha\beta}^{(\bla)} = 
                -i \int \frac{d\Omega}{2\pi i}  f(\Omega) \Tr{\left[ \vj^\alpha_{\bla} F_{\bla}^R(\Omega) \vj^\beta_{\bla} \left(  G_{\bla}^R(\Omega) - G_{\bla}^A(\Omega)  \right) - \vj^\alpha_{\bla} \left(  G_{\bla}^R(\Omega) - G_{\bla}^A(\Omega)  \right) \vj^\beta_{\bla} F_{\bla}^A(\Omega) \right]}.
                \end{equation}
Thus, Eq.~\eqref{App_B_conductivity_derivative_bare} is rewritten as
		\begin{equation}
                B_{\alpha\beta \gamma} =   -i\limvec  \partial_{\lambda_\beta} \sigma_{\alpha\gamma}^{(\bla)}.
                \end{equation}
The superscript in $\sigma^{(\bla)}$ indicates that the Green functions and current operators contain the parameter $\bla$. We also recast $B_{\alpha\beta\gamma}'$ as
        \begin{equation}
        B'_{\alpha\beta \gamma} =   - i\limvec \partial_{\lambda_\gamma} \sigma_{\alpha\beta}^{(\bla)}.
        \end{equation}
We therefore call the low-frequency divergence from $B_{\alpha\beta \gamma}$ and $B'_{\alpha\beta \gamma}$ \textit{ (static) conductivity derivative effect}. The formula is given by
        \begin{equation}
            2 \sigma^{\alpha ;\beta \gamma}_\text{sCD}
        = \frac{1}{2\omega_2}B_{\alpha\beta\gamma}+ \frac{1}{2\omega_1}{B'}_{\alpha\beta\gamma}
        = -\frac{i}{2} \limvec \left( \frac{1}{\omega_2} \partial_{\lambda_\beta} \sigma_{\alpha\gamma}^{(\bla)} + \frac{1}{\omega_1}  \partial_{\lambda_\gamma} \sigma_{\alpha\beta}^{(\bla)} \right).
        \end{equation}

Finally, we present the symmetry analysis of the anomalous NRO conductivity in Eq.~\eqref{App_second_cond_divergent}.
First, we discuss the nonreciprocal superfluid density effect in the Green function representation. Taking Eq.~\eqref{App_NRSF_third_velocity_term}, we apply the time-reversal operation $\theta$ to what is in the parenthesis. Owing to the antiunitary property of $\theta$, the expression is transformed as
		\begin{align}
                \Tr \left[\vj^{\alpha \beta \gamma}\left(G^R(\Omega)-G^A(\Omega)\right)\right] 
                        &= \Tr \left[\theta\left(G^R(\Omega)-G^A(\Omega)\right)^\dagger \theta^{-1} \theta {\vj^{\alpha \beta \gamma}}^\dagger \theta^{-1} \right],\\
                        &= \Tr \left[\left(G^R(\Omega)-G^A(\Omega)\right) \left( - \vj^{\alpha \beta \gamma} \right) \right],\\
                        &= -\Tr \left[ \vj^{\alpha \beta \gamma} \left(G^R(\Omega)-G^A(\Omega)\right) \right],\\
                        &=0.
                \end{align}
The time-reversal operation is performed in the first line and the \T{}-odd property of $\vj^{\alpha\beta\gamma}$ is implemented in the second line. As a result, the nonreciprocal superfluid density is forbidden by the \T{} symmetry. It is similarly forbidden by the \Pa{} symmetry, while the \PT{} symmetry does not make any symmetry constraint.

Following parallel calculations, we can classify $B_{\alpha\beta\gamma}$ by the \T{} and \PT{} symmetries. Imposing the \T{} symmetry, the first term in Eq.~\eqref{App_cond_deriv_one_term} is transformed as
		\begin{equation}
                \Tr \left[\vj^{\alpha \beta} F^{R} (\Omega) \vj^{\gamma} \left(G^{R}(\Omega)-G^{A}(\Omega)\right) \right]
                = - \Tr \left[ \vj^{\alpha \beta} \left(G^{R}(\Omega)-G^{A}(\Omega)\right)  \vj^{\gamma} F^{R} \right],
                \end{equation}
since the \T{}-parities of the paramagnetic and diamagnetic current operators are odd and even, respectively. Performing similar symmetry operations for other terms of $B_{\alpha\beta\gamma}$, we obtain the \T{} symmetric component given by the antisymetric part of the linear conductivity tensor
		\begin{equation}
                B_{\alpha\beta\gamma}^{(\mathcal{T})} = -\frac{i}{2} \limvec  \partial_{\lambda_\beta} \left(  \sigma_{\alpha\gamma}^{(\bla)} - \sigma_{\gamma\alpha}^{(\bla)} \right). \label{Bterm_Tsymmetric}
            \end{equation}
Since the $\sigma_{\alpha\gamma}^{(\bla)} - \sigma_{\gamma\alpha}^{(\bla)}$ is expressed by the Berry curvature,
this anomalous NRO response is given by the Berry curvature derivative.
We reproduce the expression for the Berry curvature derivative by performing the trace.
Using the eigenstates of the BdG Hamiltonian $H_\text{BdG}^{(\bla)}$, we get the relation
		\begin{align}
                        &\int \frac{d\Omega}{2 \pi i}  f(\Omega)  \Tr{\left[- \vj_\bla^{\alpha } F_{\bla}^{R} (\Omega) \vj_\bla^{\gamma} \left(G_\bla^{R}(\Omega)-G_\bla^{A}(\Omega)\right) + \vj_\bla^{\alpha } \left(G_\bla^{R}(\Omega)-G_\bla^{A}(\Omega)\right)  \vj_\bla^{\gamma} F_\bla^{R} (\Omega)  \right)}\notag \\ 
                        &= \sum_{a\neq b} \frac{f_{ab}}{E_{ab}^2} \left(\vj_\bla^{\alpha } \right)_{ab} \left(\vj_\bla^{\gamma} \right)_{ba},\\
                        &= \sum_{a\neq b} f_{a} \left[ \left(\xi^{\lambda_\alpha } \right)_{ab} \left(\xi^{\lambda_\gamma} \right)_{ba} - \left(  \alpha \leftrightarrow \gamma \right) \right],\label{App_eq1}
                \end{align}
and Eq.~\eqref{App_eq1} is recast as
		\begin{equation}
                \sum_{a\neq b} f_{a} \left[ \left(\xi^{\lambda_\alpha } \right)_{ab} \left(\xi^{\lambda_\gamma} \right)_{ba} - \left(  \alpha \leftrightarrow \gamma \right) \right] = \sum_a  - i \epsilon_{\alpha\gamma\beta} \Omega_a^\beta f_a.
                \end{equation}
Finally, we rewrite the static conductivity derivative effect for \T{} symmetric case by
        \begin{align}
            2 \sigma^{\alpha ;\beta \gamma}_\text{sCD$(\mathcal{T})$}
            &= \frac{1}{2\omega_2}B^{(\mathcal{T})}_{\alpha\beta\gamma}+ \frac{1}{2\omega_1}{B'}^{(\mathcal{T})}_{\alpha\beta\gamma},\\
            &= -\frac{i}{2} \limvec \left[ \frac{1}{\omega_2}\epsilon_{\alpha\gamma\delta} \partial_{\lambda_\beta} \left( \sum_a \Omega^{\delta}_{a} f_a  \right) + \frac{1}{\omega_1}\epsilon_{\alpha\beta\delta} \partial_{\lambda_\gamma} \left( \sum_a \Omega^{\delta}_{a} f_a  \right) \right],\\
            &= -\frac{i}{2} \left( \epsilon_{\alpha\beta\delta}  \frac{B_d^{\gamma\delta}}{\omega_1} + \epsilon_{\alpha\gamma\delta}  \frac{B_d^{\beta\delta}}{\omega_2} \right),\label{general_NRO_Berry_curvature_derivative}
        \end{align}
in the low-frequency regime.

Next, we consider the symmetry constraint of the \PT{} symmetry on the coefficient $B_{\alpha\beta\gamma}$. On the basis of the parallel discussion, the antisymmetric component is forbidden. Accordingly, $B_{\alpha\beta\gamma}$ is expressed by the symmetric component 
		\begin{equation}
                B_{\alpha\beta\gamma}^{(\mathcal{PT})} = -\frac{i}{2} \limvec  \partial_{\lambda_\beta} \left(  \sigma_{\alpha\gamma}^{(\bla)} + \sigma_{\gamma\alpha}^{(\bla)} \right),
                \end{equation}
which leads to the \textit{Drude derivative} $\hat{D_d}$ defined by
		\begin{equation}
            D_d^{\beta;\alpha\gamma}= i B_{\alpha\beta\gamma}^{(\mathcal{PT})} =  \limvec \partial_{\lambda_\beta}\left(  \int \frac{d\Omega}{2\pi} \left( - \frac{\partial f (\Omega)}{\partial \Omega}  \right) \sum_a \left(\vj_\bla^{\alpha} \right)_{aa} \left(\vj_\bla^{\gamma} \right)_{aa} 2 \Im{\left[  \left(  G_\bla^{R}(\Omega) \right)_{aa} \right]}^2  \right) .\label{App_cond_deriv_PT_sym}
                \end{equation}
This represents the intraband effect. The Drude derivative $\hat{D}_d$ vanishes in the absence of the Bogoliubov quasiparticles since it contains the Fermi-surface factor $\partial_\Omega f(\Omega)$.
Note that the parameter $\bla$ does not break the \PT{} symmetry and we can directly apply the \PT{} symmetry to Eq.~\eqref{App_B_conductivity_derivative_bare}.
The obtained result is the same as Eq.~\eqref{App_cond_deriv_PT_sym}. As a result, we obtain the static conductivity derivative effect for \PT{} symmetric case as
    \begin{align}
        2 \sigma^{\alpha ;\beta \gamma}_\text{sCD$(\mathcal{PT})$}
        &= \frac{1}{2\omega_2}B^{(\mathcal{PT})}_{\alpha\beta\gamma}+ \frac{1}{2\omega_1}{B'}^{(\mathcal{PT})}_{\alpha\beta\gamma},\\
        &= -\frac{i}{2\omega_2} D_d^{\beta;\alpha\gamma}-\frac{i}{2\omega_1} D_d^{\gamma;\alpha\beta}.
    \end{align}

\bibliographystyle{apsrev4-1}

\end{document}